\documentclass[sigconf]{acmart}

\setcopyright{none}
\copyrightyear{2026}
\acmYear{2026}
\acmDOI{XXXXXXX.XXXXXXX}

\usepackage{xcolor}
\usepackage{soul}
\usepackage{times}
\usepackage{array}
\usepackage{tabularx}
\usepackage{enumitem}
\usepackage{amsmath}
\usepackage{amsfonts}
\usepackage{multirow}
\usepackage{latexsym}
\usepackage{algorithm}
\usepackage{algorithmic}
\usepackage{makecell}

\usepackage{siunitx}
\usepackage{booktabs}
\usepackage{longtable}
\usepackage{tabularx, array, colortbl, xcolor}
\definecolor{lightgray}{gray}{0.9}
\usepackage[T1]{fontenc}

\usepackage[utf8]{inputenc}

\usepackage{microtype}

\usepackage{graphicx}
\usepackage{amsthm}
\theoremstyle{definition}
\newtheorem{definition}{Definition}

\usepackage{tcolorbox}
\tcbuselibrary{skins,breakable}
\newtcolorbox{promptbox}[1]{
  breakable,
  enhanced,
  colback=gray!1,            
  colframe=gray!40,          
  coltitle=black!90,         
  fonttitle=\bfseries\small,
  title={#1},                
  left=3pt, right=3pt, top=4pt, bottom=4pt,
  boxsep=2pt,
  arc=2pt,
  boxrule=0.5pt,
  fontupper=\ttfamily\small, 
  before skip=4pt,
  after skip=4pt
}

\newtcolorbox{dataexample*}[1]{
  breakable,
  enhanced,
  colback=cyan!8,            
  colframe=teal!60,          
  coltitle=teal!40!black,    
  fonttitle=\bfseries\small,
  title={#1},
  left=3pt, right=3pt, top=4pt, bottom=4pt,
  width=0.8\textwidth,          
  boxsep=2pt,
  arc=2pt,
  boxrule=0.5pt,
  fontupper=\ttfamily\small,
  before skip=4pt,
  after skip=4pt,
  float*=htbp  
}

\newcommand{\dataset}{{\color{black}\texttt{ComplexGraph}}}
\newcommand{\method}{{\color{black}\textsc{GraphSkill}}}

\begin{document}

\title{\textsc{GraphSkill}: Documentation-Guided Hierarchical Retrieval-Augmented Coding for Complex Graph Reasoning}
\author{Fali Wang}
\authornote{Both authors contributed equally to this research.}
\affiliation{%
  \institution{The Pennsylvania State University}
  \city{University Park}
  \state{Pennsylvania}
  \country{USA}
}
\email{fqw5095@psu.edu}

\author{Chenglin Weng}
\authornotemark[1]
\affiliation{%
  \institution{The Pennsylvania State University}
  \city{University Park}
  \state{Pennsylvania}
  \country{USA}
}
\email{cvw5844@psu.edu}


\author{Xianren Zhang}
\affiliation{%
  \institution{The Pennsylvania State University}
  \city{University Park}
  \state{Pennsylvania}
  \country{USA}
}
\email{xzz5508@psu.edu}

\author{Siyuan Hong}
\affiliation{%
  \institution{The Pennsylvania State University}
  \city{University Park}
  \state{Pennsylvania}
  \country{USA}
}
\email{sjh6636@psu.edu}

\author{Hui Liu}
\affiliation{%
  \institution{Michigan State University}
  \city{East Lansing}
  \state{Michigan}
  \country{USA}
}
\email{liuhui7@msu.edu}

\author{Suhang Wang}
\authornote{Corresponding author.}
\affiliation{%
  \institution{The Pennsylvania State University}
  \city{University Park}
  \state{Pennsylvania}
  \country{USA}
}
\email{szw494@psu.edu}

\renewcommand{\shortauthors}{xxx}

\begin{abstract}
The growing demand for automated graph algorithm reasoning has attracted increasing attention in the large language model (LLM) community. Recent LLM-based graph reasoning methods typically decouple task descriptions from graph data, generate executable code augmented by retrieval from technical documentation, and refine the code through debugging. However, we identify two key limitations in existing approaches: (i) they treat technical documentation as flat text collections and ignore its hierarchical structure, leading to noisy retrieval that degrades code generation quality; and (ii) their debugging mechanisms focus primarily on runtime errors, yet ignore more critical logical errors.
To address them, we propose {\method}, an \textit{agentic hierarchical retrieval-augmented coding framework} that exploits the document hierarchy through top-down traversal and early pruning, together with a \textit{self-debugging coding agent} that iteratively refines code using automatically generated small-scale test cases. To enable comprehensive evaluation of complex graph reasoning, we introduce a new dataset, {\dataset}, covering small-scale, large-scale, and composite graph reasoning tasks. 
Extensive experiments demonstrate that our method achieves higher task accuracy and lower inference cost compared to baselines\footnote{The code is available at \href{https://github.com/FairyFali/GraphSkill}{\textcolor{blue}{https://github.com/FairyFali/GraphSkill}}.}.
\end{abstract}



\maketitle


\section{Introduction}
Graph algorithmic reasoning tasks play an important role in many real-world applications, including social network analysis~\cite{girvan2002community}, transportation systems~\cite{hart1968formal}, and communication networks~\cite{page1999pagerank}, where solutions rely on the explicit structural and relational properties of graphs. Unlike \emph{graph learning} methods~\cite{xia2021graph,luo2024enhance,wang2024hc,yang2020nargnn}, which learn parameterized mappings from graph data distributions into labels for generalization, \emph{graph reasoning} focuses on deterministic, algorithmic inference over a given graph instance, typically involving multi-step structured operations such as traversal, search, flow computation, or compositional reasoning. Traditionally, these tasks depend on manually designed graph algorithms, which are costly to develop and difficult to adapt as problem complexity increases.

Recently, large language models (LLMs) have enabled automated graph reasoning~\cite{fatemi2023talklikegraphencoding, wang2024languagemodelssolvegraph, wang2025GraphToolInstruction, gong2025pseudocodeinjectionmagicenablingllms, zhang2024gcoderimprovinglargelanguage, li2025graphteamfacilitatinglargelanguage}, achieving substantial performance gains. LLM-based automated graph reasoning aims to replace manually implemented graph algorithms with LLMs that interpret task specifications and generate reasoning procedures (e.g., code or natural language reasoning), thereby solving graph problems automatically and reducing reliance on handcrafted algorithms. LLM-based graph reasoning methods can be broadly categorized into two classes: \emph{text-based} reasoning methods and \emph{code-based} reasoning methods. Text-based approaches serialize graph structures into natural language and leverage LLMs’ strengths in instruction following and multi-step textual reasoning through advanced prompting strategies, achieving competitive performance on small-scale and structurally simple graph tasks~\cite{fatemi2023talklikegraphencoding, wang2024languagemodelssolvegraph, chen2024graphwizinstructionfollowinglanguagemodel, luo2024graphinstruct, chai2025graphllm}. In contrast, code-based graph reasoning methods exploit LLMs’ coding capabilities to generate executable programs and delegate complex structural reasoning to program execution~\cite{cai2024codegraphenhancinggraphreasoning, zhang2024gcoderimprovinglargelanguage, li2025graphteamfacilitatinglargelanguage, wang2025GraphToolInstruction, gong2025pseudocodeinjectionmagicenablingllms}. Many code-based approaches further incorporate retrieval augmentation to compensate for limited or unreliable algorithmic knowledge in LLMs: they retrieve task-relevant algorithmic knowledge from external technical documentation (such as graph library APIs or algorithm textbooks) using flattened vector-based retrieval methods (e.g., TF-IDF~\cite{salton1988term}) and use the retrieved content to guide code generation~\cite{zhang2024gcoderimprovinglargelanguage, li2025graphteamfacilitatinglargelanguage}.

Despite their promise, existing graph reasoning methods remain insufficient for handling complex graph reasoning tasks. On the one hand, for \emph{structurally complex} large graphs, text-based reasoning methods face a major bottleneck due to context window limitations, as serializing graph structures into prompts can quickly exceed LLM context limits (e.g., a fully connected graph with approximately $180$ nodes already surpasses the $160$k-token context window of DeepSeek-V3~\cite{liu2024deepseek}). Even when large graphs fit within the context window, long-horizon reasoning over lengthy graph representations challenges LLMs’ long-context reasoning capabilities.
On the other hand, although code-based methods alleviate the context-window bottleneck by offloading structural reasoning to program execution, they face new difficulties on \textit{semantically complex} composite graph tasks, which place heavier demands on algorithmic knowledge integration and the retrieval module. Such tasks require using multiple classical graph algorithms and introduce problem-level semantic complexity (e.g., identifying strongly connected components before computing shortest paths under latency constraints in communication networks). While these composite tasks are more practical, they are also more challenging, as they require stronger algorithmic understanding and more accurate retrieval of sufficient task-relevant documents to guide code synthesis.

\begin{table}[t]
\centering
\caption{Code-Based Graph Reasoning Comparison (Ret.: retrieval; Hi.: hierarchical; Non-human: no human involvement).
}
\vskip -1em
\label{tab:method_comparison}
\resizebox{0.99\linewidth}{!}{
\begin{tabular}{lcccccc}
\hline
\textbf{Method} 
& \textbf{Coding} 
& \textbf{\makecell{Runtime\\Debug}} 
& \textbf{\makecell{Logic\\Debug}} 
& \textbf{Ret.} 
& \textbf{Hi.} 
& \textbf{\makecell{Non-\\Human}} \\
\hline
\textsc{GCoder}~\citeyearpar{zhang2024gcoderimprovinglargelanguage} 
& \checkmark &  &  & \checkmark &  & \checkmark \\
\textsc{GraphTool-Inst.}~\citeyearpar{wang2025GraphToolInstruction} 
& \checkmark &  &  &  &  & \checkmark \\
\textsc{GraphTeam}~\citeyearpar{li2025graphteamfacilitatinglargelanguage} 
& \checkmark & \checkmark &  & \checkmark &  & \checkmark \\
\textsc{CodeGraph}~\citeyearpar{cai2024codegraphenhancinggraphreasoning} 
& \checkmark &  &  &  &  &  \\
\textsc{PIE}~\citeyearpar{gong2025pseudocodeinjectionmagicenablingllms} 
& \checkmark & \checkmark & \checkmark &  &  &  \\
\textsc{Ours} 
& \checkmark & \checkmark & \checkmark & \checkmark & \checkmark & \checkmark \\
\hline
\end{tabular}
}
\vskip -2em
\end{table}

As the scalability of code-based methods for complex graph reasoning under both structural and semantic complexity, we focus on this paradigm. Tab.~\ref{tab:method_comparison} summarizes representative code-based methods and highlights two fundamental limitations of existing non-human-involved retrieval-based approaches. 
\textbf{First, existing retrieval is brittle under semantic complexity and ignores documentation hierarchy.}
For \emph{semantically complex} composite graph tasks, solving a query often requires retrieving multiple complementary algorithmic documents, placing high demands on semantic retrieval. Most RAG-style retrievers treat documentation as a flat corpus and rank top-$k$ documents by shallow similarity, which frequently misses compositional intent, leading to low recall at small $k$ and heavy noise at large $k$. Flat agentic retrieval can improve relevance via LLM screening, but it is costly and slow at scale. In contrast, technical documentation is inherently hierarchical, with coarse grouping at higher levels and precise algorithmic/API details at lower levels, making it well-suited for efficient agentic navigation. Our agentic hierarchical retrieval exploits this structure via top-down traversal and early pruning, achieving coarse-to-fine disambiguation with fewer LLM calls. It improves retrieval F1 from $\sim\!28\%$ to $79\%$ and reduces per-task search time from $23.3$s to $9.1$s compared to flat agent retrieval (Fig.~\ref{fig:retrieval_results}).
\textbf{Second, existing methods lack systematic code debugging.}
Most approaches, such as GraphTool-Instruction~\cite{wang2025GraphToolInstruction} and GCoder~\cite{zhang2024gcoderimprovinglargelanguage}, rely on one-shot code generation and thus frequently suffer from runtime and logical errors. While GraphTeam~\cite{li2025graphteamfacilitatinglargelanguage} introduces compiler-based debugging, it mainly resolves runtime issues. As shown in Fig.~\ref{fig:pilot}(a), logical errors account for most failures across models, suggesting that logic-level errors dominate yet remain unaddressed. Motivated by these observations, we pose a novel research question: \emph{How can hierarchical technical documentation be exploited to improve retrieval, and how can LLMs integrate the retrieved knowledge to generate and systematically debug robust code solutions for complex graph reasoning tasks?}

To this end, we propose {\method}, a \textbf{documentation-guided agentic hierarchical retrieval--augmented coding framework for complex graph reasoning}, consisting of a \texttt{retrieval agent} and a \texttt{coding agent}.
To retrieve high-quality documents for complex tasks at low cost, our retrieval agent exploits the hierarchical structure of technical documentation via a \textit{layer-wise agentic retrieval} algorithm over the document tree.
Starting from the root nodes, the retrieval agent traverses the document tree in a top-down manner, retaining only task-relevant nodes and pruning irrelevant branches at higher levels. It then iteratively descends to lower-level nodes until reaching the leaf-level algorithmic entries. The resulting task-relevant entries are then provided to the coding agent as external knowledge.
Guided by the retrieved documents, the coding agent integrates a \emph{self-debugging mechanism} to improve code robustness. Since debugging logical errors typically requires labeled test cases, yet task-specific test cases are rarely available in fully automated settings, the agent autonomously generates small-scale graph test cases with reliable labels. This design leverages the observation that LLMs can solve small graphs reliably when supplied with correct algorithmic guidance (Fig.~\ref{fig:pilot} shows $100\%$ accuracy on graphs with fewer than $10$ nodes in our pilot study). The agent then generates candidate code and iteratively debugs and refines it in a code compiler environment until all test cases are passed or the maximum debugging limit is reached.  The validated code is finally executed on graph instances to produce final answers.
To enable systematic evaluation of complex graph reasoning, we additionally introduce a new dataset, {\dataset}, capturing both \emph{large-scale structural complexity} and \emph{problem-level semantic complexity}, which comprises three variants: \texttt{\dataset-S} (small-scale), \texttt{\dataset-L} (large-scale), and \texttt{\dataset-C} (composite).
Experimental results show that text-based reasoning becomes nearly ineffective on our large-scale setting (e.g., DeepSeek-V3 achieves $<15\%$ accuracy), whereas code-based approaches remain effective. Moreover, {\method} consistently outperforms retrieval- and debugging-based coding baselines on composite tasks (e.g., with Qwen-2.5-7B, \textsc{GraphTeam} $56.7\%$ accuracy vs.\ \textsc{Ours} $73.3\%$), validating the benefits of our hierarchical retrieval and logic-aware self-debugging.

Our \textbf{main contributions} are: (i) We identify two key limitations in prior code-based graph reasoning, \emph{flat document retrieval} and missing \emph{logic-level debugging}; (ii) We propose {\method}, a documentation-guided agentic framework that unifies hierarchical retrieval with self-debugging code generation for complex graph reasoning; (iii) We introduce a new dataset, {\dataset}, for controlled evaluation of complex graph reasoning across increasing complexity, comprising small-scale, large-scale, and composite subsets; and (iv) Extensive experiments on \texttt{GTools} and {\dataset} show that our method improves retrieval quality and code robustness, achieving state-of-the-art performance while maintaining low inference cost compared to agentic flat retrieval.
\vskip -3em
\section{Related Work}

\noindent\textbf{Benchmarking LLMs on Graph Reasoning} Despite the success of LLMs in natural language processing, their ability to perform graph reasoning tasks, such as cycle detection, remains limited. Early benchmarks, including GraphQA \cite{fatemi2023talklikegraphencoding} and NLGraph \cite{wang2024languagemodelssolvegraph}, evaluated LLMs on basic graph reasoning and showed that, while LLMs exhibit preliminary graph reasoning capabilities, their performance falls short of ideal levels. Motivated by these findings, subsequent datasets such as GraphWiz \cite{chen2024graphwizinstructionfollowinglanguagemodel}, GraphInstruct \cite{luo2024graphinstruct}, and LLM4DyG \cite{zhang2024llm4dyg} expanded task coverage to more complex and dynamic graph reasoning settings. However, existing benchmarks remain limited to small-scale graphs and classic graph algorithmic tasks, and do not evaluate large-scale beyond the context window and composite graph reasoning. To address this gap, we introduce a new dataset, covering small- and large-scale and composite graph reasoning tasks.

\noindent\textbf{LLM-based Graph Reasoning Methods.}
Recent work has sought to improve LLMs’ limited performance on graph reasoning tasks, which can be broadly categorized into \emph{text-based} and \emph{code-based} paradigms. Text-based methods perform natural language reasoning over text-serialized graph structures using prompting strategies such as chain-of-thought and self-consistency, as studied in Talk-like-a-Graph \cite{fatemi2023talklikegraphencoding} and NLGraph \cite{wang2024languagemodelssolvegraph}, with further enhancement through instruction tuning in GraphWiz \cite{chen2024graphwizinstructionfollowinglanguagemodel} and GraphInstruct \cite{luo2024graphinstruct}. Code-based approaches generate and execute graph algorithm code, including CodeGraph \cite{cai2024codegraphenhancinggraphreasoning}, and GraphTeam \cite{li2025graphteamfacilitatinglargelanguage}. 
However, these methods remain limited by context-window constraints and, more critically, by the lack of logic-level debugging. To address these issues, we propose a self-debugging coding framework that produces robust executable programs by leveraging self-generated test cases and iterative, feedback-driven refinement. Appendix~\ref{sec:appendix_related_reasoning} for more details.
\noindent\textbf{RAG for Code-based Graph Reasoning}
Code-based methods have shown state-of-the-art performance in LLM-based graph reasoning, but approaches that rely solely on LLMs often lack explicit graph algorithm expertise, limiting code quality. Recent retrieval-augmented generation (RAG) approaches retrieve graph knowledge from technical documentation such as NetworkX~\cite{hagberg2008exploring}. For example, methods such as GraphTeam~\cite{li2025graphteamfacilitatinglargelanguage} and GCoder~\cite{zhang2024gcoderimprovinglargelanguage} treat documentation as a flat corpus and apply semantic similarity–based retrieval. Yet, these methods overlook the inherent hierarchical structure of technical documentation. 
In contrast, our approach performs agentic hierarchical retrieval over tree-structured documentation through a layer-wise agentic retrieval framework, enabling more effective and efficient retrieval. More detailed related work is discussed in Appendix~\ref{sec:appendix_related_agentic_rag}.



\section{Preliminary}
In this section, we first characterize the technical documentation structure to motivate our agentic retrieval design, then conduct pilot tests on code-based error distributions and the feasibility of self-generated test cases, and finally formalize the problem definition.

\subsection{Technical Documentation}

Technical documentation comprises structured resources written for technical users, describing reusable functions, APIs, and usage guidelines to support efficient software development. Representative examples include NumPy~\cite{harris2020array}, PyTorch~\cite{paszke2019pytorch}, and NetworkX~\cite{hagberg2008exploring}. Such documentation is typically organized in a hierarchical directory structure, arranging concepts, modules, and interfaces from coarse to fine granularity, which supports systematic navigation, improves retrieval efficiency, and reduces comprehension overhead. In coding workflows, technical documentation serves as an authoritative reference by specifying function semantics, input--output constraints, and usage examples, thereby improving implementation correctness and reliability. For graph reasoning tasks, the NetworkX~\cite{hagberg2008exploring} provides rich algorithmic interfaces that facilitate robust LLM-generated code. We formalize technical documentation as a tree structure in Sec.~\ref{sec:hierarchical_retrieval_agent}.

\subsection{Pilot Experiments}

\begin{figure}[t]
    \centering
    \includegraphics[width=\linewidth]{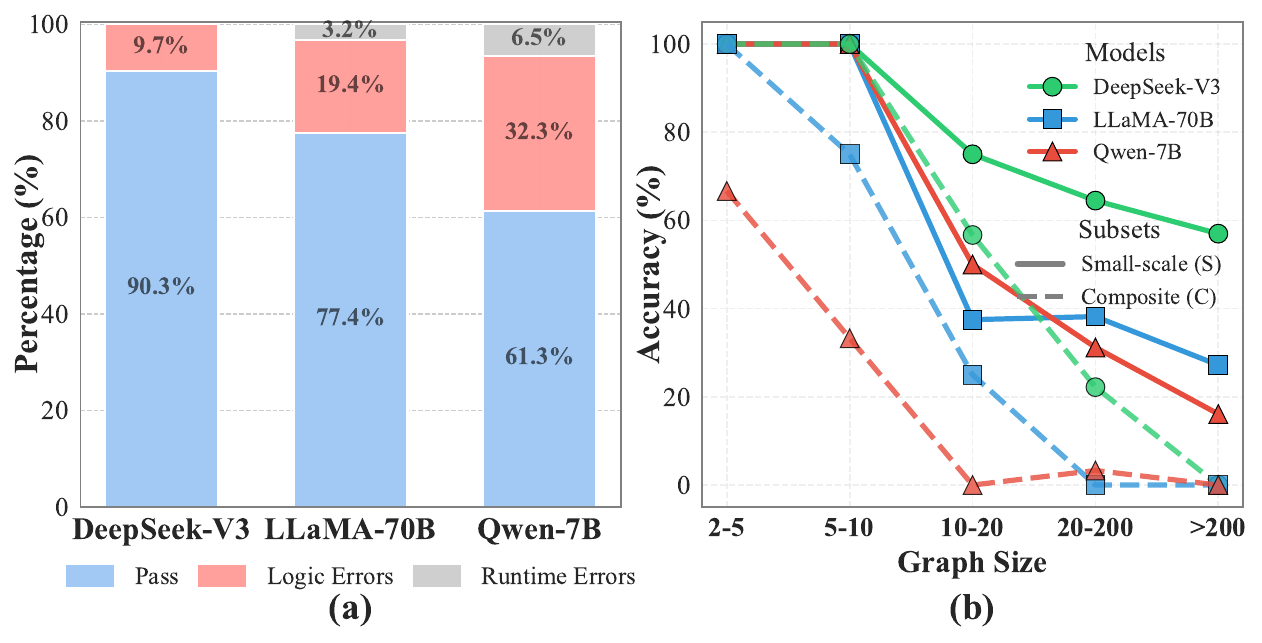}
    \vskip -1.5em
    \caption{(a) Error distribution across various LLMs. (b) Text-based reasoning accuracy (\%) across graph scales on \texttt{\dataset} using different LLMs.}
    \label{fig:pilot}
\end{figure}

\noindent\textbf{Error Distribution}.
\label{sec:appendix_debug_distribution}
Fig.~\ref{fig:pilot}(a) reports the distribution of \emph{directly passed} cases, \emph{logical errors}, and \emph{runtime errors} for zero-shot code generation on \texttt{\dataset} (detailed in Sec.~\ref{sec:our_dataset}). We define the pass rate as the fraction of cases in which the generated program passes the test suite; logical errors as cases that execute successfully but produce outputs that differ from the ground-truth answers; and the remaining cases as runtime errors. We observe that for the strongest model, DeepSeek-V3.2~\cite{liu2024deepseek}, all generated codes execute successfully, and all failures are due to logical errors, indicating that logic-level mistakes dominate. For weaker models (LLaMA-3.3-70B~\cite{dubey2024llama} and Qwen-2.5-7B~\cite{bai2023qwentechnicalreport}), fewer than $7\%$ of cases exhibit runtime errors, while most failures are logical errors ($19.4\%$ and $32.3\%$, respectively). Notably, logical errors are counted only when execution succeeds, suggesting that their true prevalence may be higher. Overall, \textit{logical errors account for the majority of code failures, motivating the need for a logic-aware debugging mechanism}.



\noindent\textbf{Feasibility of LLM-Generated Test Labels}.
\label{sec:appendix_text_reasoning_various_graph_scales}
To design an automated \emph{self-debugging} module tailored to LLM-based graph reasoning, we propose using LLMs to self-generate labeled test cases for logic-level verification and correction. To assess feasibility, we conduct a pilot study evaluating text-based reasoning on randomly generated graphs. Specifically, given a graph task, a graph instance (edge list), and input arguments, we prompt the LLM to directly output the answer via textual reasoning (prompts in Appendix~\ref{sec:appendix_prompt}). Results on the small-scale and composite subsets of {\dataset}, grouped by graph size, are reported in Fig. \ref{fig:pilot} (b). 
We observe that (i) on non-composite tasks with fewer than $10$ nodes, all three model scales (7B to 671B) achieve $100\%$ accuracy. On composite tasks, the two larger models also reach $100\%$ accuracy on graphs with fewer than $5$ nodes, and even the 7B model achieves $66.7\%$. These results suggest that LLMs can reliably label small-scale test cases, supporting automated test-case generation for self-debugging.
(ii) As graph size grows, text-based reasoning degrades sharply: for graphs with more than $200$ nodes, accuracy drops to $57\%$, $27.2\%$, and $16.1\%$ across the three models on the small-scale subset and $0$ on the composite subset. This reflects textual reasoning's limitations under long serialized inputs and long-horizon structural reasoning for large-scale graphs, motivating delegating complex reasoning to executable code.



%

\subsection{Problem Definition}
\begin{definition}[Retrieval-augmented Graph Reasoning by Coding]
\label{def:graph_reasoning_coding}
Given a task description $q$, the input consists of a set of graph instances $\mathcal{G}=\{G\}, |\mathcal{G}|\geq 1$, where each graph $G=(V,E)$ has nodes $V=\{v_i\}$ and edges $E=\{e_{ij}\}$ with $e_{ij}=(v_i,v_j,w)$, and $w$ is the weight only for weighted graphs.
An LLM denoted by $f_\theta$ is provided with a technical document collection $\mathcal{D}=\{d_i\}$ and an executable environment $\texttt{E}$. The objective is to retrieve a task-relevant subset $\mathcal{D}_q\subseteq\mathcal{D}$ and then augment the LLM to generate executable code
$c = f_\theta(q, \mathcal{D}_q)$. The code $c$ is executed on graph $G$ in environment $\texttt{E}$ to produce the answer (output)
$a = \texttt{E}(c, G), G \in \mathcal{G}.$
\end{definition}

\begin{figure}[tb]
    \centering
    \includegraphics[width=0.99\linewidth]{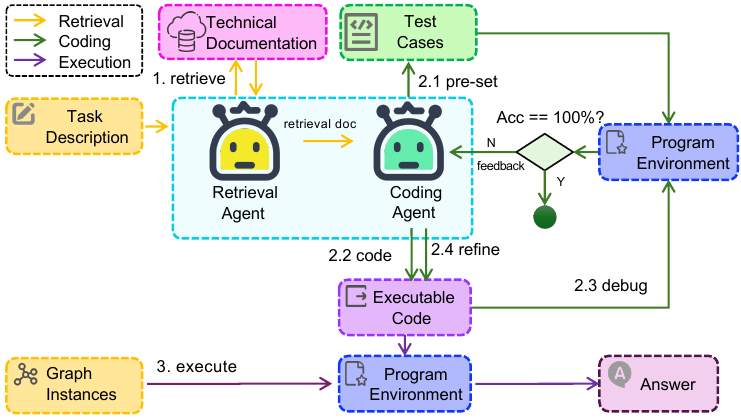}
    \vskip -1em
    \caption{Overview of the Agentic Hierarchical Retrieval-Augmented Self-debugging Coding Framework.}
    \label{fig:overall_framework}
\end{figure}

\section{Proposed Framework - {\method}}
\label{sec:method}

We propose {\method}, a technical documentation-guided agentic hierarchical retrieval--augmented self-debugging coding framework for complex graph reasoning. The framework consists of a \texttt{retrieval agent} and a \texttt{coding agent} (Fig.~~\ref{fig:overall_framework}). Specifically, given a graph reasoning task, the \texttt{retrieval agent} traverses tree-structured technical documentation to identify task-relevant algorithmic entries. Importantly, for semantically complex composite tasks, the \texttt{retrieval agent} can return a set of accurate and less-noisy complementary functions spanning multiple sub-algorithms (e.g., strongly connected components and shortest paths), enabling the \texttt{coding agent} to compose them into a code solution. To mitigate runtime and logical errors in code generation, we equip the \texttt{coding agent} with a self-debugging mechanism. It first autonomously generates labeled test cases, then produces an initial implementation, and iteratively refines the code through interaction with a compiler environment until all test cases pass or a maximum debugging budget is reached. This feedback loop is particularly critical for composite tasks, where the increased complexity often prevents one-shot code generation from being correct. Finally, the validated code is executed on the target graph instances to produce the final answers. By integrating documentation-guided agentic retrieval with compiler-driven self-debugging code generation, {\method} better supports large-scale graph reasoning than text-based reasoning, and improves composite graph reasoning compared to naive one-shot coding with flat retrieval.

\subsection{Hierarchical Retrieval Agent}
\label{sec:hierarchical_retrieval_agent}
\begin{figure}[tb]
    \centering
    \includegraphics[width=0.99\linewidth]{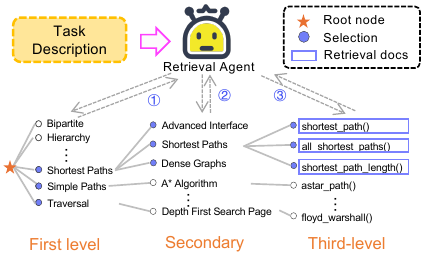}
    \vskip -1.5em
    \caption{The hierarchical retrieval agent workflow. }
    \label{fig:retrieval_agent}
\end{figure}

Prior RAG methods typically treat technical documentation as a flat text corpus and retrieve the top-$k$ documents via vector similarity such as TF-IDF. Such retrieval often relies on keywords or shallow semantic similarity, making it difficult to capture deeper task relevance; as a result, truly critical entries may not appear among the highest-ranked results. To improve recall, they often increase $k$, which introduces substantial retrieval noise and increases downstream generation cost. A straightforward alternative is to employ an LLM-based agent to filter documents using the LLM’s stronger semantic understanding, but performing document-by-document matching over large corpora incurs high LLM cost and inference latency. To address these issues, we propose a \textbf{hierarchical retrieval agent}, as shown in Fig.~\ref{fig:retrieval_agent}, that performs top-down traversal over tree-structured technical documentation, combining task-level semantic understanding with structured navigation. By pruning irrelevant branches early at higher levels, the agent identifies task-relevant entries with only a few LLM queries, enabling cost-efficient retrieval.

\noindent\textbf{Tree-Structured Technical Documentation}
We model the technical documentation $\mathcal{D}$ as a document tree $\mathcal{T}=(\mathcal{V}, \mathcal{E}, \mathcal{S})$ with maximum depth $L$. The tree is rooted at a virtual root node $v^{(0)}$ and expands level by level into $L$ hierarchical layers. The $l$th layer ($l \in \{0,\dots,L\}$) consists of a set of nodes $\mathcal{V}^{(l)}=\{v_i^{(l)}\}$, where layer $0$ is a virtual root containing a single node. Each non-leaf node $v_i^{(l)}$ is connected to its children in the next layer via edges $(v_i^{(l)}, v_j^{(l+1)}) \in \mathcal{E}$, which represent hyperlinks from the parent node web page to the child node web pages, while leaf nodes correspond to concrete document entries.
Each node $v_i^{(l)}$ can be a chapter, subsection, or individual article, and is associated with textual content ${s}_i^{(l)} \in \mathcal{S}$ that may include a chapter description or a function specification (e.g., input--output parameters, and usage examples). This formulation captures the hierarchical structure of technical documentation and provides a coarse-to-fine abstraction that supports structured navigation.

\noindent\textbf{Layer-wise Agentic Retrieval Algorithm}
Building on the hierarchical document tree $\mathcal{T}$, we introduce a \emph{Layer-wise Agentic Retrieval} algorithm, whose overall procedure is summarized in Algo.~\ref{alg:retrieval} in Appendix. The core idea is to integrate agent-driven semantic relevance assessment with structured, level-wise traversal of the document hierarchy.
Concretely, the retrieval agent $\texttt{R}$ starts from the root node $v^{(0)}$ (initialize the selected node set $\mathcal{D}_q$, Algo.~\ref{alg:retrieval}, Line 1) and examines its first-layer child nodes $\mathcal{V}^{(1)}$. Let $\mathcal{C}=\mathcal{V}^{(1)}$ denote the candidate node set at the current layer. For each node $v_i^{(1)} \in \mathcal{C}$, the agent evaluates its relevance to the task description $q$ based on the node’s semantics ${s}_i^{(1)}$. This relevance filtering operation is formalized as
\begin{equation}
\small 
\begin{aligned}
\mathcal{D}_q
&= \texttt{R}\texttt{.select\_relevant}(q, \mathcal{C}) \\
&= \{\, v_i^{(l)} \mid \texttt{R}.\texttt{is\_relevant}(q, {s}_i^{(l)}) = \texttt{Yes}, v_i^{(l)} \in \mathcal{C} \,\}
\end{aligned}
\label{eq:select_relevant}
\end{equation}
where $\mathcal{D}_q$ is updated to the set of nodes judged relevant at the current layer. The judgement prompt is shown in Appendix \ref{sec:appendix_prompt}.
The algorithm proceeds in a layer-wise manner. At a middle layer $l+1$, given the current selected set $\mathcal{D}_q \subseteq \mathcal{V}^{(l)}$, the agent collects current-layer candidate child node set
\begin{equation}
\small 
    \mathcal{C} = \{ v_j^{(l+1)} \mid v_i^{(l)} \in \mathcal{D}_q,\; (v_i^{(l)}, v_j^{(l+1)}) \in \mathcal{E} \}.
\label{eq:candidate_set}
\end{equation}
The agent then applies the same relevance selection operation in Eq.~\eqref{eq:select_relevant} to $\mathcal{C}$, yielding a new selected set $\mathcal{D}_q \subseteq \mathcal{V}^{(l+1)}$ at the current layer. This procedure is repeated iteratively for layers $l=0$ through $L-1$, until the traversal reaches the leaf level (Algo. Lines 2-5).
After the termination, the nodes remaining in $\mathcal{D}_q$ correspond to leaf nodes and thus form a set of selected technical documents. Finally, for precise selection, the retrieval agent performs a global filtering step over these candidates:
\begin{equation}
\small 
\mathcal{D}_q = \texttt{R}\texttt{.global\_filter}(q, \mathcal{D}_q),
\label{eq:global}
\end{equation}
producing the final retrieved doc subset $\mathcal{D}_q$ (Algo. Line 6). The global filtering prompt is in Appendix \ref{sec:appendix_prompt}.
Compared to static retrieval based on flattened semantic similarity, our approach employs agent-driven relevance assessment to align task semantics with both directory nodes and document entries, improving retrieval quality while avoiding excessive cost and latency by reducing the search space and pruning irrelevant branches at higher levels.

\subsection{Coding Agent with Self-Debugging}
Given the task-relevant document subset $\mathcal{D}_q$, we employ a \texttt{Coding Agent} $\texttt{C}$ to synthesize executable code for solving the graph reasoning task. As shown by the green arrows in Fig.~\ref{fig:overall_framework}, the pipeline consists of four stages: test case generation, code generation, debugging, and code refinement. The latter two stages are performed iteratively. The entire process is supported by an external executable environment $\texttt{E}$, which enables program execution and the acquisition of feedback.
Owing to the inherent complexity of certain tasks (e.g., composite tasks), the initially generated code may exhibit logical errors. To mitigate this, we introduce a \emph{self-debugging mechanism} in which the \texttt{Coding Agent} self-constructs a set of small-scale test cases with their labels, and then iteratively refines the code by validating it against these test cases until all tests pass or the debugging budget is exhausted. We describe each stage in detail below.

\noindent\textbf{Test Case Generation}
Given the task description $q$ and the retrieved documents $\mathcal{D}_q$, the \texttt{Coding Agent} exploits its reasoning capability to generate a set of small-scale graph instances (e.g., graphs with three nodes) and their corresponding labels. This design is motivated by the observation that the agent can reliably produce correct solutions on small graphs via textual reasoning (recall in Fig.\ref{fig:pilot}). Formally, this step is defined as
\begin{equation}
\small 
(\mathcal{G}_{\text{test}}, \mathcal{A}_{\text{test}})
= \texttt{C}\texttt{.generate\_test\_cases}(q), 
\label{eq:test_case_generation}
\end{equation}
where $\mathcal{G}_{\text{test}}=\{g_i\}$ is the test graphs and $\mathcal{A}_{\text{test}}=\{a_i\}$ is the labels.

\noindent\textbf{Code Generation}
The \texttt{Coding Agent} generates an initial version of executable code $c$ conditioned on the task $q$ and the retrieved documents $\mathcal{D}_q$:
\begin{equation}
c = \texttt{C}\texttt{.generate\_code}(q, \mathcal{D}_q).
\label{eq:code_generation}
\end{equation}

\noindent\textbf{Debugging and Refinement}
To evaluate code correctness, the generated code $c$ is executed in the environment $\texttt{E}$ on the test graphs $\mathcal{G}_{\text{test}}$, and the resulting outputs are compared against the reference answers $\mathcal{A}_{\text{test}}$. The test accuracy is defined as
\begin{equation}
\small 
\text{Acc}_{\text{test}}
= \frac{1}{|\mathcal{G}_{\text{test}}|}
\sum_{(g_i,a_i)\in(\mathcal{G}_{\text{test}},\mathcal{A}_{\text{test}})}
\mathbb{I}\big[\texttt{E}(c, g_i)=a_i\big],
\label{eq:code_refinement}
\end{equation}
where $\mathbb{I}[\cdot]$ denotes the indicator function.
If $\text{Acc}_{\text{test}} = 1.0$, the code $c$ is considered correct on the test cases and is returned as the final solution. Otherwise, the test cases on which the code fails, together with their expected and actual outputs, are aggregated into a feedback signal $F$, which is used by the \texttt{Coding Agent} to refine the code:
\begin{equation}
c \leftarrow \texttt{C}\texttt{.refine\_code}(q, \mathcal{D}_q, c, F).
\end{equation}
The debugging and refinement loop is executed for at most $T_{\max}$ iterations. If no code attains perfect test accuracy within this budget, we select the version with the highest $\text{Acc}_{\text{test}}$ across all iterations as the final output.
Detailed prompts are provided in Appendix~\ref{sec:appendix_prompt}.

\subsection{Code Execution on Graph Instances}
Upon obtaining a high-quality executable code $c$, we execute it in the program execution environment $\texttt{E}$ on the input graph set $\mathcal{G}=\{G_i\}_{i=1}^n$. Each graph is processed independently, yielding the outputs
\begin{equation}
\small 
\mathcal{A} = \{\, a_i \mid a_i = \texttt{E}(c, G_i),\; G_i \in \mathcal{G} \,\}.
\end{equation}
The generated code operates on graphs of arbitrary scale, constrained only by computing resources rather than LLM context size, thereby supporting both small- and large-scale graph reasoning with correctness ensured via explicit program execution. For simplicity, we omit task parameters from the notation and assume they are implicitly included in each $G_i$, as part of the task specification.

\section{Experiments}
In this section, we evaluate the proposed framework for graph reasoning on two datasets with varying graph scales and complexity, across various-size LLMs, including comparative experiments, ablations, retrieval analysis, key hyperparameter analysis, and case studies.

\begin{table*}[t]
\centering
\small 
\caption{Comparative results on \texttt{\dataset-S} and \texttt{GTools-WL} using various coding agent LLM backbones, evaluated with Accuracy. The best result in each column is bolded (except for CodeGraph and PIE, whose coding processes involve human intervention). The Avg.\ column reports the mean over all values. Baselines: (i) Text-based Reasoning; (ii) Code-based (No Retrieval, No Debugging); (iii) Code-based (Retrieval, No Debugging); (iv) Code-based (Retrieval + Debugging); and (v) Code-based (Human-involved).}
\label{tab:main_results_S}
\vskip -1.2em
\resizebox{0.82\linewidth}{!}{
\renewcommand{\arraystretch}{1.1}
\begin{tabular}{llccccccc}
\hline
 &  & \multicolumn{3}{c}{\texttt{\dataset-S}} & \multicolumn{3}{c}{\texttt{GTools-WL}} & \textbf{Avg.} \\
\cmidrule(lr){3-5} \cmidrule(lr){6-8}
 & \textbf{Method} 
& \textbf{DeepSeek} & \textbf{LLaMA-70B} & \textbf{Qwen-7B}
& \textbf{DeepSeek} & \textbf{LLaMA-70B} & \textbf{Qwen-7B}
&  \\
\hline
\multirow{4}{*}{(i)}
 & \textsc{Zero-shot} & 67.3 & 32.3 & 20.9 & 68.0 & 43.5 & 40.5 & 45.4 \\
 & \textsc{Few-shot} & 64.6 & 35.0 & 16.4 & 78.5 & 52.0 & 45.5 & 48.7 \\
 & \textsc{CoT} & 62.3 & 32.7 & 20.9 & 79.0 & 45.5 & 29.5 & 45.0 \\
 & \textsc{Bag-CoT} & 66.4 & 24.6 & 15.5 & 69.5 & 36.0 & 26.5 & 39.8 \\
\hline
\multirow{2}{*}{(ii)}
 & \textsc{Zero-shot Coding} & 81.8 & 73.2 & 45.4 & 88.0 & 82.5 & 50.0 & 70.1 \\
 & \textsc{Few-shot Coding} & 83.6 & 69.1 & 58.6 & 90.0 & 82.5 & 58.0 & 73.6 \\
\hline
\multirow{2}{*}{(iii)}
 & \textsc{TFIDF+Coding} & 94.1 & 90.0 & 45.0 & 80.0 & 77.5 & 57.5 & 74.0 \\
 & \textsc{SentBERT+Coding} & 94.1 & 80.5 & 50.0 & 85.0 & 75.0 & 45.0 & 71.6 \\
\hline
\multirow{3}{*}{(iv)}
 & \textsc{GraphTeam} & 82.7 & 79.1 & 64.6 & 85.0 & 75.0 & 70.0 & 76.1 \\
 & \textsc{TFIDF+CodeAgent} & 94.6 & 90.5 & 76.8 & 90.0 & 90.0 & 70.0 & 85.3 \\
 & \textsc{SentBERT+CodeAgent} & 93.7 & 92.3 & 87.7 & 90.0 & 90.0 & 70.0 & 87.3 \\
\hline
\multirow{2}{*}{(v)}
 & \textsc{CodeGraph} & 94.6 & 94.6 & 85.9 & 90.0 & 95.0 & 80.0 & 90.0  \\
 & \textsc{PIE} & 94.2 & 94.6 & 87.2 & 84.4 & 90.0 & 80.0 & 88.4 \\
\hline
\rowcolor{lightgray}
 & \textsc{Ours} & \textbf{96.4} & \textbf{96.4} & \textbf{91.8} & \textbf{100.0} & \textbf{100.0} & \textbf{73.5} & \textbf{93.0} \\
\hline
\end{tabular}}
\end{table*}

\begin{table*}[t]
\centering
\small
\caption{Comparative results on \texttt{\dataset-L} and \texttt{GTools-EL} using various coding agent LLM backbones, evaluated with the Accuracy metric. Best in each column is bolded except for CodeGraph and PIE. (i)–(v) follow the same statements as above Tab. \ref{tab:main_results_S}.}
\label{tab:main_results_L}
\vskip -1.2em
\resizebox{0.82\linewidth}{!}{
\renewcommand{\arraystretch}{1.1}
\begin{tabular}{llccccccc}
\hline
 &  & \multicolumn{3}{c}{\texttt{\dataset-L}} & \multicolumn{3}{c}{\texttt{GTools-EL}} & \textbf{Avg.} \\
\cmidrule(lr){3-5} \cmidrule(lr){6-8}
 & \textbf{Method} 
& \textbf{DeepSeek} & \textbf{LLaMA-70B} & \textbf{Qwen-7B}
& \textbf{DeepSeek} & \textbf{LLaMA-70B} & \textbf{Qwen-7B}
&  \\
\hline
\multirow{4}{*}{(i)}
 & \textsc{Zero-shot} & 12.4 & 3.8 & 0.0 & 78.9 & 42.2 & 41.1 & 29.7 \\
 & \textsc{Few-shot} & 13.3 & 2.4 & 0.0 & 83.3 & 62.8 & 48.3 & 35.0 \\
 & \textsc{CoT} & 11.9 & 5.7 & 0.0 & 84.4 & 53.9 & 46.1 & 33.7 \\
 & \textsc{Bag-CoT} & 8.6 & 3.3 & 0.0 & 71.7 & 33.3 & 24.4 & 23.6 \\
\hline
\multirow{2}{*}{(ii)}
 & \textsc{Zero-shot Coding} & 85.2 & 66.2 & 39.1 & 93.3 & 94.4 & 48.3 & 71.1 \\
 & \textsc{Few-shot Coding} & 84.7 & 44.8 & 48.1 & 83.3 & 91.7 & 64.4 & 69.5 \\
\hline
\multirow{2}{*}{(iii)}
 & \textsc{TFIDF+Coding} & 94.3 & 89.5 & 47.6 & 80.6 & 72.2 & 55.6 & 73.3 \\
 & \textsc{SentBERT+Coding} & 94.3 & 88.1 & 70.5 & 91.7 & 77.8 & 36.1 & 76.4 \\
\hline
\multirow{3}{*}{(iv)}
 & \textsc{GraphTeam} & 87.6 & 81.0 & 70.5 & 94.4 & 75.0 & 77.8 & 81.1 \\
 & \textsc{TFIDF+CodeAgent} & 94.3 & 94.3 & 80.0 & 83.3 & 91.7 & 72.2 & 85.9 \\
 & \textsc{SentBERT+CodeAgent} & \textbf{99.1} & \textbf{99.1} & 97.6 & 94.4 & 94.4 & 83.3 & 94.7 \\
\hline
\multirow{2}{*}{(v)}
 & \textsc{CodeGraph} & 99.1 & 99.1 & 90.0 & 100.0 & 94.4 & 67.8 & 91.7  \\
 & \textsc{PIE} & 97.1 & 96.7 & 90.0 & 94.4 & 100.0 & 77.8 & 92.7 \\
\hline
\rowcolor{lightgray}
 & \textsc{Ours} & \textbf{99.1} & \textbf{99.1} & \textbf{99.1} & \textbf{100.0} & \textbf{100.0} & \textbf{86.1} & \textbf{97.2} \\
\hline
\end{tabular}
}
\vskip -1em
\end{table*}


\subsection{Experimental Setup}

\noindent\textbf{Datasets, Metrics, and LLM Backbones.} We conduct experiments on both \texttt{GTools}~\cite{wang2025GraphToolInstruction} (including two subsets: \texttt{GTools-EL} for large graphs and \texttt{GTools-WL} for small graphs) and the proposed {\dataset} datasets below, using NetworkX~\cite{hagberg2008exploring} as the technical documentation. Accuracy is used as the primary evaluation metric. DeepSeek-V3.2~\cite{liu2024deepseek} is consistently employed as the retrieval agent, while the coding agent is selected from DeepSeek-V3.2, LLaMA-3.3-70B~\cite{dubey2024llama}, and Qwen-2.5-7B~\cite{bai2023qwentechnicalreport}. The temperature is fixed at $0.1$ all the time. Additional details are provided in Appendix~\ref{sec:appendix_experiment_setup}.

\noindent\textbf{Baselines.}
Baselines are grouped into five categories.
(i) \textbf{Pure text-based reasoning} methods include \textsc{Zero-shot}, \textsc{Few-shot}, \textsc{Chain-of-Thought} (\textsc{CoT}), and \textsc{Build-a-Graph} (\textsc{BaG-CoT})~\cite{wang2024languagemodelssolvegraph}.
Code-based methods are further categorized by the use of retrieval and debugging:
(ii) \textbf{No retrieval or debugging}, including \textsc{Zero-shot Coding}, which directly prompts the LLM to generate executable code, and \textsc{Few-shot Coding}, which provides a small number of task-irrelevant example code solutions as in-context demonstrations;
(iii) \textbf{Retrieval without debugging}, including \textsc{TFIDF+ Coding} and \textsc{SentBERT+Coding}, which augment \textsc{Zero-shot Coding} with document retrieval using TF-IDF \cite{salton1988term} and Sentence-BERT~\cite{reimers2019sentence}, respectively, but do not perform iterative code correction.
(iv) \textbf{Retrieval with debugging}, including \textsc{GraphTeam}~\cite{li2025graphteamfacilitatinglargelanguage}, which leverages LlamaIndex~\cite{llamaindexweb} for flattened semantic retrieval and incorporates runtime debugging.
We additionally include \textsc{TFIDF+ CodeAgent} and \textsc{SentBERT+CodeAgent}, which use TF-IDF and Sentence BERT to retrieve top-$k$ documents from flattened documentation and employ the same coding agent with self-debugging as our method.
(v) \textbf{Human-involved} baselines include \textsc{CodeGraph}~\cite{cai2024codegraphenhancinggraphreasoning}, which relies on manually selected documents, and \textsc{PIE}~\cite{gong2025pseudocodeinjectionmagicenablingllms}, which depends on manually provided pseudocodes and labeled test cases for debugging.
Baseline details are provided in Appendix~\ref{appendix:baselines}.

\subsection{The Proposed {\dataset} Dataset}
We introduce a new synthetic benchmark, {\dataset}, for evaluating \emph{complex} graph reasoning under both \emph{structural} and \emph{semantic} complexity. {\dataset} contains three subsets: small-scale \texttt{\dataset-S} with small graphs (3--200 nodes), large-scale \texttt{\dataset-L} with large graphs (5k and 10k nodes) to stress scalability beyond context-window limits, and composite \texttt{\dataset-C} consisting of composite tasks that require composing two or three primitive graph algorithms via sequential, parallel, or conditional structures. All graph instances are generated with NetworkX \cite{hagberg2008exploring}, and labels are produced by executing human-crafted reference implementations. Detailed task lists, construction procedures, and statistics are provided in Appendix \ref{sec:our_dataset}. From the results, existing ``large-graph'' benchmarks (e.g., \texttt{GTools-EL}, $\leq 200$ nodes) remain within typical LLMs' context windows and thus do not challenge text-based reasoning (Tab. \ref{tab:main_results_L}). In contrast, on our truly large-scale graphs, text reasoning degrades sharply, demonstrating that our {\dataset} introduces meaningful structural complexity. For composite tasks, semantic complexity reveals clear limitations of prior retrieval-plus-coding baselines (e.g., \textsc{GraphTeam} in Tab.~\ref{tab:main_results_C}), while human-curated retrieval and test-case debugging yield much larger gains, highlighting deficiencies in existing retrieval/coding components and validating the necessity of our proposed composite dataset (detailed analysis in Appendix \ref{appendix:analysis_our_dataset}).

\begin{table}[t]
\centering
\small
\caption{Comparative results on \texttt{\dataset-C}.} 
\label{tab:main_results_C}
\vskip -1em
\resizebox{0.95\linewidth}{!}{
\renewcommand{\arraystretch}{1.1}
\begin{tabular}{llccc}
\hline
 & \textbf{Method} 
 & \textbf{DeepSeek} 
 & \textbf{LLaMA-70B} 
 & \textbf{Qwen-7B} \\
\hline
\multirow{4}{*}{(i)}
 & \textsc{Zero-shot}   & 43.3 & 5.6 & 4.4 \\
 & \textsc{Few-shot}    & 43.3 & 4.4 & 2.2 \\
 & \textsc{CoT}         & 42.2 & 2.2 & 3.3 \\
 & \textsc{Bag-CoT}     & 55.6 & 0.0 & 11.1 \\
\hline
\multirow{2}{*}{(ii)}
 & \textsc{Zero-shot Coding} & 64.4 & 56.7 & 6.7 \\
 & \textsc{Few-shot Coding}  & 78.9 & 51.1 & 10.0 \\
\hline
\multirow{2}{*}{(iii)}
 & \textsc{TFIDF+Coding}     & 73.3 & 66.7 & 36.7 \\
 & \textsc{SentBERT+Coding}  & 73.3 & 66.7 & 55.6 \\
\hline
\multirow{3}{*}{(iv)}
 & \textsc{GraphTeam}        & 76.7 & 70.0 & 55.6 \\
 & \textsc{TFIDF+CodeAgent}  & 84.4 & 88.9 & 56.7 \\
 & \textsc{SentBERT+CodeAgent} & 73.3 & 88.9 & 66.7 \\
\hline
\multirow{2}{*}{(v)}
 & \textsc{CodeGraph}        & 95.6 & 88.9 & 66.7 \\
 & \textsc{PIE}              & 100.0 & 92.2 &  70.0 \\
\hline
\rowcolor{lightgray}
 & \textsc{Ours}             & \textbf{95.6} & \textbf{96.7} & \textbf{73.3} \\
\hline
\end{tabular}
}
\vskip -1em
\end{table}


\subsection{Results and Analysis} 
Tables~\ref{tab:main_results_S} and~\ref{tab:main_results_L} compare our method with a range of baselines on both \textit{small-scale} and \textit{large-scale} graph data from two datasets, evaluated across multiple coding-agent LLM backbones. Several key observations emerge.
(1) Our method achieves the best performance on both data scales, with average accuracies of $93.0\%$ (small-scale) and $97.2\%$ (large-scale). Among methods without human involvement, it significantly outperforms the strongest baseline, SentBERT+CodeAgent, by $5.7\%$ and $2.5\%$ on small- and large-scale graphs, respectively. Moreover, our method matches or exceeds human-involved approaches (CodeGraph and PIE) in most settings, demonstrating strong effectiveness for automated graph reasoning.
(2) Text-based reasoning baselines are consistently inferior to coding-based methods, with the gap especially pronounced on large-scale graphs. For example, Zero-shot text reasoning achieves only $33.7\%$ accuracy on large-scale datasets, compared to $71.1\%$ for Zero-shot coding. This gap highlights the limitations of pure text-based reasoning for large-scale and structurally complex graph tasks, stemming from the inherent difficulty LLMs face in long-horizon, precise structural reasoning, whereas coding-based approaches mitigate this limitation by reasoning at the algorithmic level and delegating complex execution to programs.
(3) Comparing categories (ii) and (iii), which differ only in the use of retrieval augmentation, we observe consistent performance gains from retrieval; for example, under the Zero-shot coding setting, incorporating TFIDF-based retrieval improves accuracy by $2.9\%$ on small-scale graphs and $2.2\%$ on large-scale graphs, indicating that retrieval provides more relevant contextual information for effective code generation.
(4) Further improvements arise when incorporating explicit debugging. Compared to category (iii), methods in category (iv) integrate self-debugging and achieve substantial gains; for instance, TFIDF+Coding improves from $74.0\%$ to $85.3\%$ on small-scale datasets and from $73.3\%$ to $85.9\%$ on large-scale datasets after applying debugging, confirming the effect of leveraging debugging feedback during coding.
(5) Within category (iv), GraphTeam underperforms TFIDF+CodeAgent and SentBERT+CodeAgent because it relies primarily on runtime checks (such as runtime and semantic errors) and lacks labeled test-case-based logical verification. This limitation results in performance gaps of up to $11.2\%$ on small-scale graphs and $13.6\%$ on large-scale graphs, underscoring the importance of logic-level debugging guided by test cases.


Tab.~\ref{tab:main_results_C} reports results on \emph{composite} graph reasoning tasks, which introduce additional semantic complexity at the problem level. The overall trends remain consistent with previous results, with our method achieving the best performance. Notably, small models struggle more on this subset: Qwen-2.5-7B attains an accuracy of $73.3\%$, substantially lower than DeepSeek-V3.2 ($95.6\%$) and LLaMA-3.3-70B ($96.7\%$). This gap suggests that composite graph reasoning requires stronger semantic task understanding, where larger models demonstrate clear advantages, thereby challenging the effectiveness of small language models on complex graph reasoning tasks.

\subsection{Ablation Study}
Tab.~\ref{tab:ablation_study} reports ablation results on \texttt{\dataset} across different LLM backbones to assess the impact of each component within the {\method}. We compare the full {\method} with five ablated variants: \textsc{w/o Retrieval}, \textsc{w/o Hierarchy} (using TF-IDF filtering instead of hierarchical retrieval), \textsc{w/o Debugging}, \textsc{w/o Test Cases} (debugging without test-case assistance), and \textsc{w/o Deb + Ret}.  
From the results, we observe that  
(i) both retrieval and self-debugging are critical. Removing retrieval or debugging reduces average accuracy by $13.4\%$ and $5.0\%$, respectively, while removing both leads to a larger drop of $19\%$, indicating strong complementarity;  
(ii) removing hierarchical traversal in retrieval alone causes a $8.7\%$ performance degradation, highlighting its effect in mitigating noisy retrieval;  
(iii) excluding self-generated test cases in debugging results in an average $4.0\%$ accuracy drop, confirming their role in correcting code errors.  
Notably, for larger models (DeepSeek and LLaMA-3-70B), which have few runtime code errors, removing debugging yields a performance drop equal to removing test cases, indicating that their debugging gains primarily stem from labeled test-case–based logical error correction, whereas smaller models (Qwen-2.5-7B) additionally benefit from runtime execution checks.



\begin{table}[t]
\centering
\caption{Ablation study on \texttt{\dataset}.}
\vskip -1em
\label{tab:ablation_study}
\resizebox{0.77\linewidth}{!}{
\begin{tabular}{lcccc}
\hline
\textbf{Setting} & \textbf{DeepSeek} & \textbf{LLaMA} & \textbf{Qwen} & \textbf{Avg.} \\
\hline
\textsc{Full}             & 96.0 & 96.6 & 82.6 & 91.7 \\
\textsc{w/o Retrieval}    & 91.9 & 81.8 & 61.2 & 78.3 \\
\textsc{w/o Hierarchy}    & 92.5 & 87.9 & 68.5 & 83.0 \\
\textsc{w/o Debugging}    & 94.2 & 96.0 & 69.9 & 86.7 \\
\textsc{w/o Test cases}   & 94.2 & 96.0 & 73.0 & 87.7 \\
\textsc{w/o Ret + Deb}    & 86.0 & 80.8 & 51.3 & 72.7 \\
\hline
\end{tabular}
}
\vskip 0.5em
{\footnotesize 
\textit{Note:} All values are averaged over the small-scale and composite subsets.
}
\end{table}

\subsection{Retrieval Performance Analysis} 
\begin{figure}[t]
    \centering
    \includegraphics[width=0.9\linewidth]{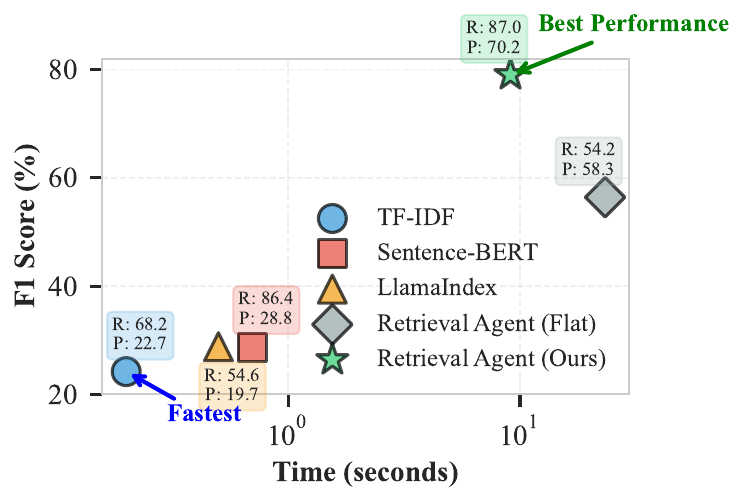}
    \vskip -1.5em
    \caption{Retrieval performance and time on small-scale subset.}
    \label{fig:retrieval_results}
\end{figure}

Fig.~\ref{fig:retrieval_results} and \ref{fig:retrieval_results_c} compare our retrieval method with retrieval baselines on \texttt{\dataset}, evaluating Recall, Precision, F1, and average retrieval time per graph problem (see Appendix~\ref{sec:appendix_metric} for metrics).  
We observe that (i) our method consistently outperforms all baselines in retrieval quality, with especially large gains in precision, improving from the strongest vector-based baseline $28.8$ to $70.2$, while maintaining high recall. This demonstrates our method’s effectiveness in reducing noisy retrieval results without sacrificing coverage.  
(ii) These improvements come with increased retrieval cost: vector-based methods complete in under $1$s on average, whereas our method requires approximately $9$s, reflecting a clear performance--efficiency trade-off. We argue that this overhead is acceptable for code-based graph reasoning, where a single generated program can be reused across many graph instances.  
(iii) Compared with a flat retrieval agent that retrieves an initial set of documents via TF-IDF and examines them, our hierarchical retrieval improves performance by enabling agent-guided branch selection to reduce noisy candidates, followed by precise document-level decisions; early branch pruning further narrows the search space and reduces retrieval time.

\subsection{Hyperparameter Analysis}
\begin{figure}[b]
    \centering
    \vskip -1.5em
    \includegraphics[width=0.8\linewidth]{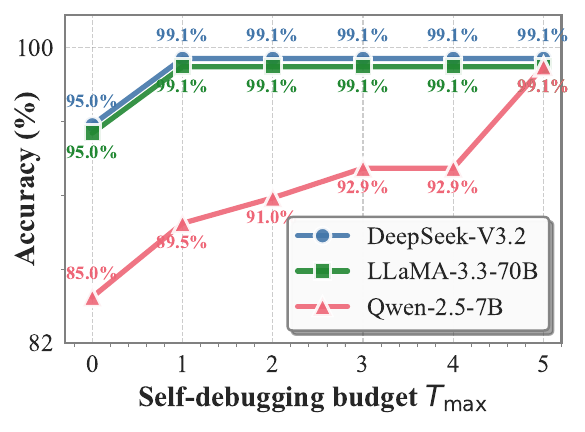}
    \vskip -1em
    \caption{Sensitivity analysis of the hyper-parameter w.r.t. self-debugging budget $T_{\max}$ on \texttt{\dataset-L}. }
    \label{fig:hyper_correction}
\end{figure}

Fig.~\ref{fig:hyper_correction} studies the effect of the self-debugging budget $T_{\max}$, which control the number of code updates guided by test-case feedback. We vary $T_{\max}$ in $\{0, 1,2,3,4,5\}$. The results show that:
(i) Increasing $T_{\max}$ from 0 to 1 yields a clear and consistent improvement in code generation quality for both large and small models, confirming the effectiveness of self-debugging.
(ii) The small model (Qwen-7B) improves steadily as $T_{\max}$ increases, from $85\%$ to $99\%$ when $T_{\max}$ grows from $1$ to $5$. In contrast, large models (DeepSeek-V3.2 and LLaMA-70B) benefit mainly from a single self-debugging step, with additional budget yielding negligible gains. This suggests that large models typically converge after one correction, whereas small models require multiple iterations to refine their code. 
We attribute this difference to variations in coding capability.
(iii) With sufficient self-debugging budget, small models can reach performance comparable to large models, consistent with prior findings on test-time scaling that increased inference-time compute can substantially improve model performance and, in some cases, enable small language models to match or even outperform larger models~\cite{brown2024large, wu2025inference, snell2025scaling, wang2025agenttts,wang2025generalizing}.

\subsection{Inference Cost}
Tab.~\ref{tab:inference_cost} compares the per-task inference cost of our method with traditional vector-retrieval-based approaches (TF-IDF and Sentence-BERT) combined with coding, as well as \textsc{GraphTeam}. We decompose the total inference cost into retrieval cost and coding cost. The results show that:
(i) TFIDF- and Sentence-BERT--based methods incur negligible retrieval cost but exhibit the highest code generation cost (approximately \$\(9\times10^{-4}\) per task), as they require a large number of documents to ensure sufficient coverage, substantially increasing the input cost during coding.
(ii) \textsc{GraphTeam} adopts higher-quality vector retrieval via LlamaIndex \cite{Liu_LlamaIndex_2022} with the OpenAI \texttt{text-embedding-ada-002} model \cite{textembeddingada}, which reduces the code generation cost (approximately \$\(5\times10^{-4}\)). However, the overhead from index construction incurs a substantial upfront cost.
(iii) Our method achieves the lowest total inference cost (around \$\(2\times10^{-4}\)). Although it incurs a non-negligible retrieval cost, the precise selection of task-relevant documents significantly lowers the code generation cost, effectively amortizing the retrieval overhead.

\begin{table}[t]
\centering
\caption{Breakdown of Inference Costs per Task (DeepSeek-V3).} 
\label{tab:inference_cost}
\vskip -1.1em
\resizebox{0.8\linewidth}{!}{
\begin{tabular}{l@{\hskip 15pt}r@{\hskip 15pt}r@{\hskip 15pt}r}
\toprule
\textbf{Method} 
& \textbf{Retrieval} 
& \textbf{Coding} 
& \textbf{Total} \\
\midrule
\textsc{TFIDF+Coding}    
& $\sim$0    
& 9     
& 9     \\
\textsc{SentBERT+Coding} 
& $\sim$0  
& 9     
& 9     \\
\textsc{GraphTeam}       
& 800$^{\dagger}$ / 0.1 
& 5     
& ${\sim}$5 \\
\textsc{Ours}            
& 2     
& 2     
& ${\sim}$4   \\
\bottomrule
\end{tabular}
}
\vskip 0.5em
{\footnotesize 
\textit{Note:} All costs in units of $10^{-4}$ USD.
$^{\dagger}$Index building cost (one-time, amortized).
}
\vskip -1em
\end{table}


\begin{figure}[h]
    \centering
    \vskip -1em
    \includegraphics[width=\linewidth]{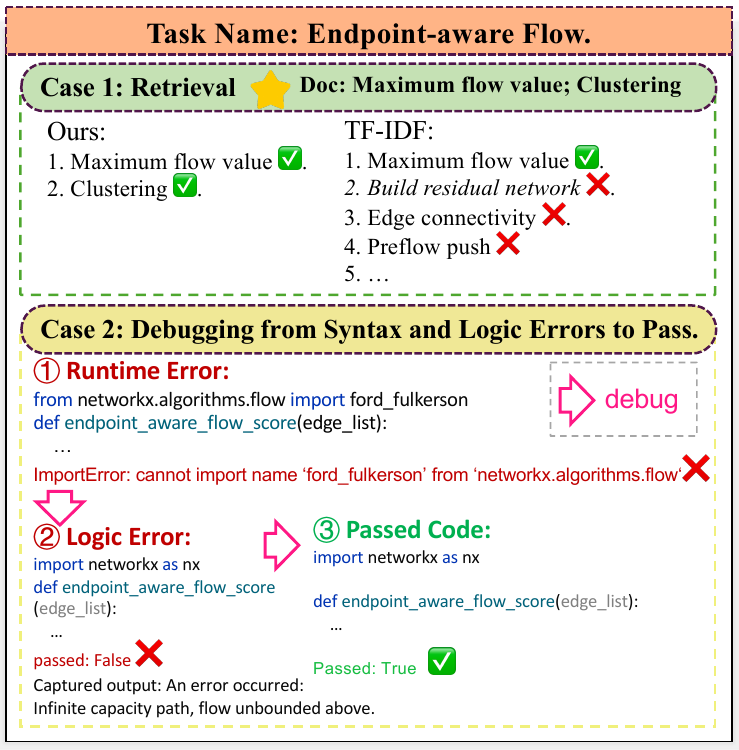}
    \vskip -1em
    \caption{Case Study: Retrieval and Debugging.}
    \label{fig:case}
\end{figure}

\subsection{Case Study}
Fig.~\ref{fig:case} presents two cases on the \texttt{Endpoint-aware Flow} task with LLaMA-3-70B. In Case 1, TF-IDF returns only one relevant document alongside substantial noise, whereas our method retrieves all required documents with no irrelevant entries. In Case 2, our self-debugging mechanism fixes both runtime and logical errors in the generated code, producing a functionally correct implementation.

\section{Conclusion}
In this study, we identify two core limitations of existing code-based graph reasoning: flat retrieval that is insufficient for handling complex tasks and the lack of debugging mechanisms that address logic-level errors. To address these limitations, we propose {\method}, a documentation-guided framework that couples hierarchical, agentic retrieval with test-case-driven self-debugging code generation. We also introduce {\dataset}, a benchmark spanning small-scale, large-scale, and composite graph reasoning. Across \texttt{GTools} and {\dataset}, {\method} consistently improves retrieval quality and task accuracy without incurring much inference cost or latency, with the largest gains on composite tasks.





\bibliographystyle{ACM-Reference-Format}
\bibliography{sample-base}

\appendix



\label{sec:appendix}
\section{Appendix}
\label{sec:appendix}



\subsection{Prompt Design} 
\label{sec:appendix_prompt}

\begin{promptbox}{Prompt for Relevance Judgment in Eq.\eqref{eq:select_relevant}}
\noindent
You will be provided with a description of a graph task and a dictionary of chapter names along with their descriptions.

\textbf{User Query}
\begin{verbatim}
{query}
\end{verbatim}

\textbf{Instructions}
\begin{enumerate}
    \item Determine the most appropriate \texttt{NetworkX} function required to solve the task.
    \item Based on the selected function, identify the relevant documentation chapters.
\end{enumerate}

\textbf{Available Chapters (key $\rightarrow$ description)}
\begin{verbatim}
{chapter_dict}
\end{verbatim}

\noindent
Choose {unique} items from \textbf{Available Chapters} that are essential for the given task.

\textbf{Constraints}
\begin{itemize}
    \item Keys {must match exactly} and are case-sensitive (use the dictionary keys as-is).
    \item If fewer than \texttt{\{top\_k\}} eligible keys exist, return as many as exist ($\geq 0$).
\end{itemize}

\textbf{Selection Guidelines (apply in order)}
\begin{enumerate}
    \item {Direct capability}: Prefer chapters whose descriptions indicate functions, algorithms, or APIs that can directly fulfill the query.
    \item {Specificity over breadth}: Prefer narrowly scoped chapters targeting the requested operation.
    \item {Disambiguation}: When multiple chapters are similar, choose the one most likely to lead to an immediately usable function.
\end{enumerate}

\textbf{Output Format (Strict)}
\begin{itemize}
    \item Return \textbf{only} a Python list literal using double quotes, e.g.,
\end{itemize}

\begin{verbatim}
["key1", "key2", "key3"]
\end{verbatim}

\begin{itemize}
    \item No explanations, no code fences, and no extra text.
    \item If nothing fits, return \texttt{[]}.
\end{itemize}

\paragraph{Internal Validation}
\begin{itemize}
    \item All returned items must be present in the {Available Chapters} keys.
    \item All items must be unique.
\end{itemize}
\end{promptbox}

\begin{promptbox}{Prompt for Global Filtering in Eq.\eqref{eq:global}}
Determine if the function is helpful for addressing the query. Think about the graph type and read carefully the User Query.

Return only one word: Yes or No. If uncertain, answer No. No explanations.

\end{promptbox}
\begin{promptbox}{Prompt for Test Case Generation in Eq.\eqref{eq:test_case_generation}}
{You are a test case generation agent.}

Generate test cases for evaluating Python programs that solve a graph algorithm task.

\textbf{Constraints:}
\begin{itemize}
  \item Do \emph{not} solve the task.
  \item Do \emph{not} include algorithms, reasoning, or hints.
  \item Use \emph{only} the task description and graph specification.
  \item All test cases must be valid and deterministic.
\end{itemize}

\textbf{Requirements:} Each test case must include:
\begin{enumerate}
  \item A concrete graph instance
  \item All required input parameters
  \item The expected output
\end{enumerate}

\textbf{Output format:}
\begin{itemize}
  \item Return a Python-compatible list of dictionaries with keys \texttt{"input"} and \texttt{"expected\_output"}.
  \item Do \emph{not} include explanations or extra text.
\end{itemize}

\textbf{Task description:} \texttt{\{task\_description\}}

\textbf{Graph specification:} \texttt{\{graph\_type\_description\}}

\end{promptbox}
\begin{promptbox}{Prompt for Code Generation in Eq.\eqref{eq:code_generation}}
\noindent
You are tasked with implementing a Python function to solve a graph problem.

\textbf{Relevant Documentation}
The following \texttt{NetworkX} documentation may be helpful for solving this problem:
\begin{verbatim}
{docs_section}
\end{verbatim}

\textbf{Task Description}
\begin{verbatim}
{question_text}
\end{verbatim}

\textbf{Graph Properties}
\begin{itemize}
    \item {Type}: \texttt{\{directed\_text\}}, \texttt{\{weighted\_text\}}
\end{itemize}

\textbf{Input}
\begin{itemize}
    \item \texttt{edge\_list}: A \texttt{\{weighted\_text\}} \texttt{\{directed\_text\}} graph represented as a list of edges.
    \begin{itemize}
        \item If weighted: each edge is \texttt{[source, target, weight]}, where \texttt{weight} is a floating-point value.
        \item If unweighted: each edge is \texttt{[source, target]}.
    \end{itemize}
    \item \texttt{\{args\_desc\}}
\end{itemize}

\textbf{Requirements}
\begin{enumerate}
    \item Implement a function that takes an edge-list representation of the graph.
    \item Convert the edge list into a \texttt{NetworkX} \texttt{\{nx\_graph\_class\}} object.
    \item Use \texttt{NetworkX} functions to solve the task.
    \item Return the result as specified in the task description.
    \item Return \texttt{None} if any error occurs during execution.
\end{enumerate}

\textbf{Important Notes}
\begin{itemize}
    \item Carefully read the provided \texttt{NetworkX} documentation.
    \item Pay close attention to parameter names, input types, and return values.
    \item Handle edge cases such as empty graphs and disconnected components.
    \item Use \texttt{try--except} blocks to catch errors and return \texttt{None} on failure.
\end{itemize}

\textbf{Test Case (for Validation)}
\noindent
\textbf{Input:}
\begin{verbatim}
{Edge List}
{Arguments}
\end{verbatim}

\noindent
\textbf{Expected Output:}
\texttt{\{Answer\}}

\medskip
\noindent
Test your function using the input above and ensure that it produces the expected output. Print the output to verify correctness.

\textbf{Your Task}
Generate a complete Python function that solves this problem. The implementation should include:
\begin{enumerate}
    \item Necessary imports (e.g., \texttt{import networkx as nx})
    \item A clear and well-defined function signature
    \item Conversion from edge list to a \texttt{NetworkX} graph
    \item Appropriate \texttt{NetworkX} operations to solve the task
    \item A return statement with the correct result type
    \item Robust error handling (returning \texttt{None} on failure)
\end{enumerate}

\noindent
Provide the implementation in a Python code block.

\end{promptbox}

\begin{promptbox}{Prompt for Code Refinement in Eq.\eqref{eq:code_refinement}}
\noindent
The following code failed to produce the correct output for a graph problem.

\textbf{Original Task:}

\texttt{\{original\_query\}}

\textbf{Failed Code}

\begin{verbatim}
{error_code}
\end{verbatim}

\textbf{}{Error or Incorrect Output}

\texttt{\{error\_output\}}

\textbf{Test Case}

\textbf{Input:}
\begin{verbatim}
{test_input_str}
\end{verbatim}

\noindent
\textbf{Expected Output:}
\texttt{\{Answer\}}

\textbf{Instructions}
Analyze the error and provide a corrected version of the code that:
\begin{enumerate}
    \item Fixes any syntax or runtime errors
    \item Produces the correct output matching the expected result
    \item Handles edge cases properly
    \item Returns \texttt{None} if any error occurs
    \item Tests the corrected implementation using the provided input and prints the output
\end{enumerate}

\noindent
Provide the corrected implementation in a Python code block.

\end{promptbox}







\begin{promptbox}{\footnotesize Retrieval Prompt Variants (Clustering Coefficient Task)}

\textbf{Task Name (Formal)}  
Given a weighted, undirected graph $G=\{V, E\}$, compute the clustering coefficient of a vertex $v$, which measures the tendency of nodes to form clusters.

\textbf{Task Name (Alternative)}  
For a weighted graph $G=\{V, E\}$, determine the clustering coefficient of node $v$, indicating how many of its neighbors are mutually connected.

\textbf{No Task Name (Implicit Intent)}  
Given a weighted, undirected graph $G=\{V, E\}$, compute a coefficient for node $v$ that quantifies the connectivity among its neighbors.

\textbf{Real-World (Context 1)}  
In a social network where nodes represent people and edge weights denote interaction frequency, compute the local coefficient for person $v$ to measure how tightly their friends are connected.

\textbf{Real-World (Context 2)}  
In an interbank lending network where nodes represent banks and weights indicate exposures, compute local and global coefficients to assess how often a bank’s counterparties are interconnected.

\end{promptbox}

\begin{promptbox}{Prompt for Zero-shot Text-based Reasoning Baseline}
\noindent\texttt{\{task\_description\}}\\
The graph information \texttt{\{graph\_info\}}.\\
Please answer following \texttt{\{answer\_template\}} {WITHOUT any explanation}.\\
The output should be a \texttt{\{output\_type\}}.\\
You must return the correct answer, reasoning {CAREFULLY}.

\end{promptbox}
\begin{promptbox}{Prompt for Few-shot Text-based Reasoning Baseline}
\noindent\texttt{\{task\_description\}}\\
You can check the related info to help you reason about the answer:\\
\texttt{\{prompt\_hint\}}\\
The graph information \texttt{\{graph\_info\}}.\\
Please answer following \texttt{\{answer\_template\}} {WITHOUT any explanation}.\\
The output should be a \texttt{\{output\_type\}}.\\
You must return the correct answer, reasoning {CAREFULLY}.

\end{promptbox}

\begin{promptbox}{Prompt for Zero-shot Code-based Reasoning Baseline}
\noindent
Given the following task description:
\begin{verbatim}
{question_text}
\end{verbatim}

\noindent
Generate a Python function that solves this task for a \texttt{\{weighted\_text\}} \texttt{\{directed\_text\}} graph.

\textbf{Input}
\begin{itemize}
    \item \texttt{edge\_list}: A \texttt{\{weighted\_text\}} \texttt{\{directed\_text\}} graph represented as a list of edges.
    \begin{itemize}
        \item If weighted: each edge is \texttt{[source, target, weight]}, where \texttt{weight} is a floating-point value.
        \item If unweighted: each edge is \texttt{[source, target]}.
        \item \texttt{\{args\_desc\}}
    \end{itemize}
\end{itemize}

\textbf{Output}
\begin{itemize}
    \item Store the final answer in a variable named \texttt{result}.
    \item The result must match the expected return type for the given task.
\end{itemize}

\noindent
Provide the Python implementation below.
\end{promptbox}

\begin{promptbox}{Prompt for Few-shot Code-based Reasoning Baseline}

\noindent
Given the task description:
\begin{verbatim}
{question_text}
\end{verbatim}

\noindent
Generate a Python function that solves this task for a \texttt{\{weighted\_text\}} \texttt{\{directed\_text\}} graph.

\textbf{Input}
\begin{itemize}
    \item \texttt{edge\_list}: A \texttt{\{weighted\_text\}} \texttt{\{directed\_text\}} graph represented as a list of edges.
    \begin{itemize}
        \item If weighted: each edge is \texttt{[source, target, weight]}, where \texttt{weight} is a floating-point value.
        \item If unweighted: each edge is \texttt{[source, target]}.
        \item \texttt{\{args\_desc\}}
    \end{itemize}
\end{itemize}

\textbf{Output}
\begin{itemize}
    \item Store the final answer in a variable named \texttt{result}.
    \item The result must match the expected return type for the given task.
\end{itemize}

\textbf{Example}
\begin{verbatim}
{few_shot_examples}
\end{verbatim}

\noindent
Now, solve the task described above following a similar approach.

\textbf{Your Code:}

\end{promptbox}

\subsection{Our Algorithm}
Our proposed algorithm is in Algo. \ref{alg:retrieval}.

\begin{algorithm}[t]
\caption{Layer-wise Agentic Retrieval}
\label{alg:retrieval}
\small 
\begin{algorithmic}[1]
\REQUIRE Task $q$; document tree $\mathcal{T}$ with depth $L$, Retrieval Agent \texttt{R}
\ENSURE Task-relevant document subset $\mathcal{D}_q$

\STATE Initialize $\mathcal{D}_q \leftarrow \{v^{(0)}\}$


\FOR{$l = 0$ \TO $L-1$}
    \STATE $\mathcal{C} \leftarrow \{\, v_j^{(l+1)} \mid v_i^{(l)} \in \mathcal{D}_q  ,\,  (v_i^{(l)}, v_j^{(l+1)}) \in \mathcal{E} \}$ 
    \STATE $\mathcal{D}_q \leftarrow \texttt{R.select\_relevant}(q, \mathcal{C})$ \hfill $\triangleright$ \text{\texttt{See Eq.\eqref{eq:select_relevant}.}}
\ENDFOR

\STATE $\mathcal{D}_q \leftarrow \texttt{R.global\_filter}(q, \mathcal{D}_q)$
\RETURN $\mathcal{D}_q$
\end{algorithmic}
\end{algorithm}

\subsection{Our Proposed {\dataset} Dataset}
\label{sec:our_dataset}
\noindent\textbf{\texttt{\dataset-L(S)}: A Large(Small)-Scale Graph Reasoning Dataset.}
Prior work has proposed many graph reasoning datasets, but they primarily focus on \emph{small-scale graphs}. For example, GraphQA \cite{fatemi2023talklikegraphencoding} and NLGraph \cite{wang2024languagemodelssolvegraph} typically involve graphs with $3$--$10$ nodes, which can be accommodated by existing LLMs. For example, Qwen-7B \cite{bai2023qwentechnicalreport} with an $8$k context window supports complete graphs of $\sim40$ nodes, and DeepSeek-V3 \cite{liu2024deepseek} with a $160$k window supports $\sim180$ nodes. Such settings do not capture the challenges of \emph{large-scale graph reasoning} under long-context abilities and window size limit.
To address this gap, we construct {\dataset}, a synthetic dataset spanning graph sizes from $3$ to $10$k nodes, including a small-scale subset \texttt{\dataset-S} (graphs with $3$--$200$ nodes) and a large-scale subset \texttt{\dataset-L} (graphs with $5$k and $10$k nodes). Following \cite{wu-etal-2025-grapheval36k}, we select {$14$} classical graph algorithm tasks (see Tables~\ref{tab:small_scale_task_overview}, \ref{tab:task_overview} in Appendix). Task descriptions are written by graph-domain experts manually, and graph instances are generated using NetworkX \cite{hagberg2008exploring}.
{\dataset} covers diverse graph types (directed or undirected, weighted or unweighted, cyclic or acyclic, connected or disconnected, and sparse or dense). For each task, we generate only semantically appropriate graph types; for instance, connected component counting is evaluated exclusively on unweighted undirected graphs, as the weights do not affect the result. Edge weights for directed graphs are sampled uniformly from $[0.0, 10.0]$, and task-specific inputs (e.g., source--target pairs) are selected uniformly at random. Labels are generated using human-crafted reference code. A complete example is provided at the end. 

\noindent\textbf{\texttt{\dataset-C}: A Composite Graph Reasoning Dataset.}
Existing datasets do not provide support for \emph{complex graph reasoning tasks}. To address this gap, we introduce \texttt{\dataset-C}, a dataset dedicated to \emph{composite graph tasks}, in which multiple classical graph algorithms are composed into a single task that requires multi-step reasoning. For instance, a composite task may require first performing a topological sort to establish precedence constraints, and subsequently counting the number of paths from a source node to a target node within the reachable subgraph that satisfy these constraints.
The construction of \texttt{\dataset-C} consists of three stages.
(1) \textbf{Subtask selection}: We randomly sample two or three atomic classical graph algorithm tasks from \texttt{\dataset-S} as candidate subtasks.
(2) \textbf{Task composition and description generation}: The selected subtasks are combined according to a valid task logic. Specifically, given algorithms A and B (or A, B, C), we adopt one of three composition patterns: \emph{sequential composition}, where algorithm A is executed first and its output serves as an intermediate result for algorithm B; \emph{parallel composition}, where algorithms A and B are executed concurrently and their outputs are merged via an aggregation operator; and \emph{conditional composition}, where algorithm A is executed first and, depending on whether its output satisfies a predefined condition, either algorithm B or algorithm C is subsequently executed. Only compositions that are semantically and logically coherent are retained; otherwise, the corresponding task instances are discarded. Natural language descriptions of composite tasks are generated using GPT-5, with the prompt template provided below.
(3) \textbf{Graph instance and label generation}: Graph instances are directly reused from \texttt{\dataset-S}, while ground-truth labels are obtained by executing human-crafted reference implementations.
Detailed statistics of the composite tasks are reported in Table~\ref{tab:composite_tasks}. A complete example is provided at the end of the Appendix. Each composite task is described in detail below: 

\noindent\textbf{Sequential Composition Tasks: }

\begin{enumerate}[nosep,leftmargin=*]
\item \textbf{Clustering Coefficient \& Shortest Path.}  
Given an unweighted undirected graph $G$ and two nodes $s$ and $t$:  
(1) compute one shortest path $P$ from $s$ to $t$;  
(2) for each node $v \in P$, compute its local clustering coefficient;  
(3) identify the node on $P$ with the highest clustering coefficient, breaking ties by choosing the node closer to $s$ along the path.  
The output consists of the selected node together with the clustering coefficients of all nodes on the path.

\item \textbf{Strongly Connected Components \& Diameter.}  
Given an directed graph $G$:  
(1) compute all strongly connected components (SCCs) of $G$;  
(2) for each SCC, construct the induced subgraph and treat it as an undirected graph;  
(3) compute the diameter of each induced subgraph.  
The output is the list of diameters for all SCCs.

\item \textbf{Strongly Connected Components \& Maximum Flow.}  
Given a weighted directed graph $G$ and a fixed sink node $t$:  
(1) compute the strongly connected components of $G$;  
(2) for each SCC $C$, select a representative source node $s_C$ (e.g., the node with the smallest ID in $C$);  
(3) compute the maximum flow from $s_C$ to $t$;  
(4) compute the local clustering coefficient of $s_C$ in the undirected projection of $G$;  
(5) define a score $S_C = \mathrm{flow}(s_C,t)\cdot(1+\mathrm{clustering}(s_C))$.  
The output is the SCC with the highest score $S_C$.
\end{enumerate}

\noindent\textbf{Parallel Composition Tasks:}

\begin{enumerate}[nosep,leftmargin=*]
\setcounter{enumi}{3}
\item \textbf{Pair Tightness Score.}  
Given a weighted undirected graph $G$ and two nodes $s$ and $t$:  
(1) compute the weighted shortest-path distance $d(s,t)$;  
(2) compute the number of common neighbors of $s$ and $t$;  
(3) compute the local clustering coefficients $C(s)$ and $C(t)$.  
These quantities are combined into a tightness score
$T(s,t)=\frac{|N_{\mathrm{common}}|+\frac{C(s)+C(t)}{2}}{1+d(s,t)}.$
The output is the score $T(s,t)$.

\item \textbf{Shortest Path \& Clustering Coefficient (Bridge Hub Identification).}  
Given an undirected graph $G$ and an integer $k$:  
(1) compute the local clustering coefficient $C(v)$ for each node $v$;  
(2) compute the average shortest-path distance $\bar{d}(v)$ from $v$ to all other nodes;  
(3) define a bridge-hub score $S(v)=(1-C(v))/\bar{d}(v)$.  
The output is the top-$k$ nodes ranked by $S(v)$ together with their scores.

\item \textbf{Maximum Flow \& Clustering Coefficient.}  
Given a weighted directed graph $G$, a source node $s$, and a sink node $t$:  
(1) compute the maximum flow value $F$ from $s$ to $t$;  
(2) compute the local clustering coefficients $C(s)$ and $C(t)$ in the undirected projection of $G$.  
These results are combined into an endpoint-aware flow score
$S = F \cdot \left(1 + \frac{C(s)+C(t)}{2}\right).$
The output is the score $S$.
\end{enumerate}

\noindent\textbf{Conditional Composition Tasks}

\begin{enumerate}[nosep,leftmargin=*]
\setcounter{enumi}{6}
\item \textbf{Eulerian Path \& Diameter.}  
Given an unweighted undirected graph $G$ and two nodes $s$ and $t$:  
(1) check whether $G$ admits an Eulerian path;  
(2) if an Eulerian path exists, compute and return the diameter of $G$;  
(3) otherwise, compute the local clustering coefficients of $s$ and $t$ and return the larger one.

\item \textbf{Graph Connectivity \& Component Diameter.}  
Given an unweighted undirected graph $G$:  
(1) check whether $G$ is connected;  
(2) if $G$ is connected, compute and return its diameter;  
(3) otherwise, identify the largest connected component (by number of nodes) and compute the diameter of that component only.

\item \textbf{Strongly Connected Components \& Eulerian Path Check.}  
Given an unweighted directed graph $G$:  
(1) check whether $G$ has an Eulerian path;  
(2) if an Eulerian path exists, return the number of nodes in the largest strongly connected component;  
(3) otherwise, return the total number of SCCs.
\end{enumerate}

\begin{promptbox}{Prompt for Composite Task Generation}
You are given a set of graph algorithms.
Compose them into a single composite graph task following one of the allowed composition patterns (sequential, parallel, or conditional).

\textbf{Constraints}:

- Output only the final task description.
- Do not include explanations, algorithms, or solution steps.
- Do not reveal intermediate results or execution order.
- Ensure the task is logically consistent and self-contained.
- Clearly specify the required graph type and all necessary inputs.
- Define a single, well-formed output.
\end{promptbox}

\subsection{Necessity of the Proposed {\dataset}: An Empirical Analysis}
\label{appendix:analysis_our_dataset}
From Tab.~\ref{tab:main_results_L}, we observe that although \texttt{GTools-EL} is labeled as a large-graph benchmark, its graphs contain no more than $200$ nodes and thus remain well within the context windows of mainstream LLMs. As a result, this setting places limited pressure on text-based reasoning in terms of context capacity and long-horizon structural inference. First, text-based reasoning does not clearly break down on \texttt{GTools-EL}, and in some cases achieves performance comparable to coding-based methods (e.g., Qwen-7B attains similar accuracy under few-shot text prompting and zero-shot coding). Second, the overall results on \texttt{GTools-WL} and \texttt{GTools-EL} are broadly similar, further suggesting that existing ``large-graph'' settings are insufficient to meaningfully separate reasoning paradigms by scalability. In contrast, under our truly large-scale graph setting, text-based reasoning degrades drastically: even the strongest model (DeepSeek) achieves only around $10\%$ accuracy, while a 7B model is nearly unable to solve the tasks (close to $0\%$). This stark contrast indicates that {\dataset} introduces substantial structural complexity that poses a genuine challenge to existing text-based reasoning approaches and better exposes their scalability limits.

Moreover, Tab.~\ref{tab:main_results_C} highlights another critical dimension: \emph{problem-level semantic complexity} in composite tasks substantially amplifies difficulties in both retrieval and code generation, making naïve retrieval-plus-coding pipelines unreliable. For instance, a representative coding-based baseline, \textsc{GraphTeam}, attains only the suboptimal performance on this setting (DeepSeek: $76.7\%$ vs.\ our $95.6\%$), indicating that prior methods do not adequately address two key challenges. First, composite tasks require retrieving complementary knowledge across multiple sub-algorithms, which raises the bar for retrieval accuracy and coverage. Second, semantic complexity introduces more risks on logic-level failures, which demand logic-level debugging beyond runtime checks. Controlled comparisons further support this conclusion: holding the coding agent fixed, merely changing the retrieval strategy (\textsc{TFIDF+CodeAgent} vs.\ \textsc{Ours}) yields large gains, $11.2\%$ on DeepSeek and $16.6\%$ on Qwen. These results show that semantic complexity in composite tasks poses a substantial retrieval challenge to existing coding-based graph reasoning methods. These observations indicates that existing retrieval and coding components are insufficiently robust to semantic complexity, and supports the necessity of our composite dataset.

\subsection{Detailed Experimental Setup: Dataset, Metric, and Baseline}
\label{sec:appendix_experiment_setup}

\subsubsection{Datasets}
We conduct experiments on two datasets: both \texttt{GTools}~\cite{wang2025GraphToolInstruction} and our proposed {\dataset}. Below, we describe \texttt{GTools}; {\dataset} is introduced in detail above.


\noindent\textbf{\texttt{GTools}}~\cite{wang2025GraphToolInstruction} is a synthetic graph dataset generated by controlling the number of nodes and the probability of edge formation between node pairs. It provides two variants based on graph scale: WL-Graph (within-limited graphs) and EL-Graph (exceed-limited graphs).
We summarize the dataset statistics based on its GitHub repository and use the off-the-shelf test set for evaluation, as in Table \ref{tab:gtools_wl_el_stats}. 
\texttt{GTools} test set covers $11$ graph reasoning tasks.
Each task contains $20$ graph instances, with $10$ directed and $10$ undirected graphs. The original purpose of \texttt{GTools} is to evaluate the ability of LLMs to invoke graph-related tools. Thus, each instance comprises a task description, graph instance, and labeled tool names, tool parameters, and outputs. We use only the task description, graph instance, and output.
\begin{table}[htbp]
\centering
\small
\caption{Statistics of the GTools dataset. WL denotes small-scale graphs and EL denotes large-scale graphs.}
\label{tab:gtools_wl_el_stats}
\renewcommand{\arraystretch}{1.15}
\resizebox{0.7\linewidth}{!}{
\begin{tabular}{lccc}
\toprule
\textbf{Task Name} & \textbf{Instances} & \textbf{Nodes} & \textbf{Edges} \\
\midrule
\multicolumn{4}{c}{\textbf{GTools-WL (Small-scale)}} \\
\midrule
Maximum Flow              & 20 & 12--39 & 34--150  \\
Shortest Path Length      & 20 & 10--38 & 37--148  \\
Topological Sort          & 10 & 5--37  & 7--265   \\
Node Existence Check      & 20 & 12--40 & 38--300  \\
Edge Existence Check      & 20 & 6--38  & 19--280  \\
Cycle Detection           & 20 & 14--40 & 13--400  \\
Number of Edges           & 20 & 5--40  & 10--274  \\
Number of Nodes           & 20 & 8--40  & 17--269  \\
Path Existence Check      & 20 & 16--39 & 31--277  \\
Node Degree               & 20 & 11--40 & 32--259  \\
Clique Enumeration        & 10 & 11--38 & 24--149  \\
\midrule
\textbf{Total (WL)}        & \textbf{200} & 5--40 & 7--400 \\
\midrule
\multicolumn{4}{c}{\textbf{GTools-EL (Large-scale)}} \\
\midrule
Maximum Flow              & 20 & 11--51  & 17--627   \\
Shortest Path Length      & 20 & 10--88  & 51--728   \\
Node Existence Check      & 20 & 12--66  & 56--921   \\
Edge Existence Check      & 20 & 12--71  & 36--988   \\
Cycle Detection           & 20 & 12--99  & 11--952   \\
Number of Edges           & 20 & 5--71   & 14--818   \\
Number of Nodes           & 20 & 6--62   & 20--803   \\
Path Existence Check      & 20 & 12--97  & 42--1345  \\
Node Degree               & 20 & 13--127 & 53--900   \\
\midrule
\textbf{Total (EL)}        & \textbf{180} & 5--127 & 11--1345 \\
\bottomrule
\end{tabular}
}
\end{table}



\subsubsection{Metrics}
\label{sec:appendix_metric}
\noindent\textbf{Accuracy for graph reasoning.}
We adopt \emph{accuracy} as the evaluation metric, which measures the proportion of test instances for which the model produces correct predictions. Formally, given a test set $\mathcal{D}=\{(q_i, \mathcal{G}_i, \mathcal{A}_i)\}_{i=1}^{N}$ consisting of $N$ tasks, where $q_i$ denotes the task query, $\mathcal{G}_i$ the corresponding set of graph instances, and $\mathcal{A}_i$ the set of ground-truth answers, and given the model predictions $\hat{\mathcal{A}}_i$, the accuracy is defined as
\begin{equation}
\mathrm{Accuracy} = \frac{1}{N} \sum_{i=1}^{N} \frac{1}{|\mathcal{A}_i|} \sum_{j=1}^{|\mathcal{A}_i|} \mathbb{I}\big(\hat{\mathcal{A}}_{ij} = \mathcal{A}_{ij}\big),
\end{equation}
where $\mathbb{I}(\cdot)$ is the indicator function that equals $1$ if the predicted answer $\hat{\mathcal{A}}_{ij}$ matches the ground-truth answer $\mathcal{A}_{ij}$, and $0$ otherwise.

\noindent\textbf{Recall Metric for Retrieval.}
Let $\mathcal{D}$ denote the set of graph tasks, with $|\mathcal{D}| = N$. For each task $t \in \mathcal{D}$, let $\mathcal{R}_t$ be the set of ground-truth required documents, and let $\mathcal{H}_t$ denote the set of documents retrieved by the model. The per-task recall is defined as
\begin{equation}
\mathrm{Recall}_t =
\begin{cases}
\dfrac{|\mathcal{H}_t \cap \mathcal{R}_t|}{|\mathcal{R}_t|}, & \text{if } |\mathcal{R}_t| > 0, \\[6pt]
1, & \text{if } |\mathcal{R}_t| = 0,
\end{cases}
\end{equation}
which measures the fraction of required functions that are successfully retrieved for task $t$.
The overall recall is computed as the average of per-task recall scores across all tasks:
\begin{equation}
\mathrm{Recall} = \frac{1}{N} \sum_{t \in \mathcal{D}} \mathrm{Recall}_t.
\end{equation}

\noindent\textbf{Precision Metric for Retrieval.}
For each task $t \in \mathcal{D}$, the per-task precision is defined as
\begin{equation}
\mathrm{Precision}_t =
\begin{cases}
\dfrac{|\mathcal{H}_t \cap \mathcal{R}_t|}{|\mathcal{H}_t|}, & \text{if } |\mathcal{H}_t| > 0, \\[6pt]
0, & \text{if } |\mathcal{H}_t| = 0,
\end{cases}
\end{equation}
which measures the fraction of retrieved documents that are relevant for task $t$.
The overall precision is computed as the average of per-task precision scores across all tasks:
\begin{equation}
\mathrm{Precision} = \frac{1}{N} \sum_{t \in \mathcal{D}} \mathrm{Precision}_t.
\end{equation}

\noindent\textbf{F1 Metric for Retrieval.}
For each task $t \in \mathcal{D}$, the per-task F1 score is defined as the harmonic mean of per-task precision and recall:
\begin{equation}
\mathrm{F1}_t =
\begin{cases}
\dfrac{2 \cdot \mathrm{Precision}_t \cdot \mathrm{Recall}_t}{\mathrm{Precision}_t + \mathrm{Recall}_t}, & \text{if } \mathrm{Precision}_t + \mathrm{Recall}_t > 0, \\[6pt]
0, & \text{otherwise}.
\end{cases}
\end{equation}
The overall F1 score is computed as the average of per-task F1 scores across all tasks:
\begin{equation}
\mathrm{F1} = \frac{1}{N} \sum_{t \in \mathcal{D}} \mathrm{F1}_t.
\end{equation}

\begin{figure}[t]
    \centering
    \includegraphics[width=0.8\linewidth]{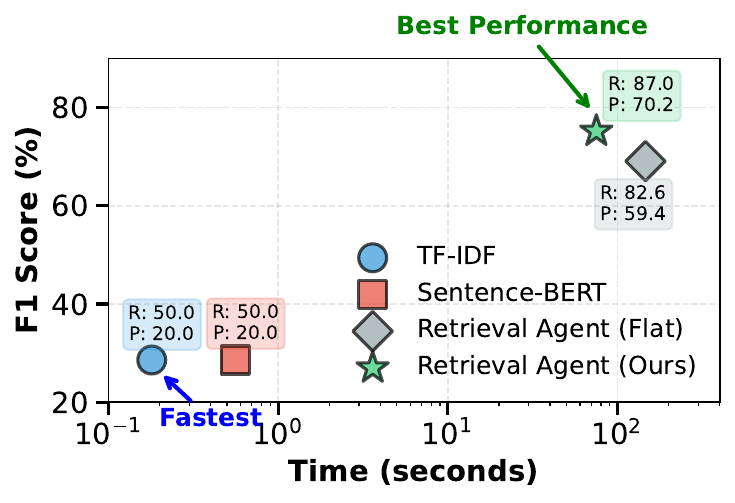}
    \vskip -1.5em
    \caption{Retrieval performance and time on composite subset.}
    \label{fig:retrieval_results_c}
\end{figure}

\subsubsection{Baselines}
\label{appendix:baselines}
We categorize the baselines into two main groups: {text-based reasoning} and {code-based reasoning}. Due to the diversity of code-based approaches, we further divide them into four subcategories: {code-based reasoning without retrieval or debugging}, {code-based reasoning with retrieval but without debugging}, {code-based reasoning with both retrieval and debugging}, and {human-involved code-based reasoning}. All baselines use NetworkX~\cite{hagberg2008exploring} as the technical documentation if retrieval is included. We detail each category and the corresponding methods below.

\paragraph{(i) Text-based reasoning}
Following Talk-like-a-Graph~\cite{fatemi2023talklikegraphencoding}, we consider several representative text-based reasoning baselines, including zero-shot prompting (\textsc{Zero-shot}), few-shot in-context learning (\textsc{Few-shot})~\cite{brown2020language}, chain-of-thought prompting (\textsc{CoT})~\cite{wei2022chain}, and the build-a-graph prompt (\textsc{CoT-BAG})~\cite{wang2024languagemodelssolvegraph}. Specifically,
\begin{itemize}[nosep,leftmargin=*]
    \item \textsc{Zero-shot}: The model is provided only with the task description and is asked to directly generate the textual reasoning output, without any task-specific demonstrations.
    \item \textsc{Few-shot}: A small number of input–output exemplars are included in the prompt to guide the model in performing the task on new inputs.
    \item \textsc{CoT}: The prompt includes step-by-step reasoning exemplars, enabling the model to generate intermediate reasoning chains when solving tasks.
    \item \textsc{CoT-BAG}: This method augments the task description with an explicit instruction, by appending ``Let us construct a graph with the nodes and edges first'', to encourage the model to reason over the underlying graph structure explicitly.
\end{itemize}
The detailed prompts are provided in Sec.~\ref{sec:appendix_prompt}.

\paragraph{(ii) Code-based reasoning without retrieval or debugging}
This category includes baselines that solve graph reasoning tasks by generating executable code, without employing retrieval augmentation or debugging mechanisms. These methods assess the inherent code-generation capability of LLMs when applied to graph reasoning tasks. Accordingly, we consider both zero-shot and few-shot coding settings, namely \textsc{Zero-shot Coding} and \textsc{Few-shot Coding}. Specifically,
\begin{itemize}[nosep, leftmargin=*]
    \item \textsc{Zero-shot Coding}: The model is provided only with the task description and is required to directly generate executable code, without any additional guidance or augmentation.
    \item \textsc{Few-shot Coding}: A small number of input–output exemplars are included in the prompt to guide the model in generating code-based solutions for graph reasoning.
\end{itemize}
The detailed prompts are provided in Sec.~\ref{sec:appendix_prompt}.

\paragraph{(iii) Code-based reasoning with retrieval but without debugging}
This category comprises baselines that address graph reasoning tasks by generating executable code with retrieval augmentation, but without explicit debugging mechanisms. Methods in this category include \textsc{TFIDF+Coding}, and \textsc{SentBERT+Coding}. Specifically,
\begin{itemize}[nosep, leftmargin=*]
    \item \textsc{TFIDF+Coding}: This retrieves relevant documentation using TF-IDF indexing~\cite{salton1988term} and conditions the LLM on the retrieved top-k results for code generation.
    \item \textsc{SentBERT+Coding}: This performs retrieval using Sentence-BERT embeddings~\cite{reimers2019sentence} to obtain semantically relevant top-k documentation, which is then used to guide code generation.
\end{itemize}

\paragraph{(iv) Code-based reasoning with both retrieval and debugging}
This category includes baselines that solve graph reasoning tasks by generating executable code with both retrieval augmentation and debugging mechanisms. Representative methods in this category include \textsc{GraphTeam}~\cite{li2025graphteamfacilitatinglargelanguage}, \textsc{TFIDF+CodeAgent}, and \textsc{SentBERT+CodeAgent}. Specifically,
\begin{itemize}[nosep, leftmargin=*]
    \item \textsc{GraphTeam}~\cite{li2025graphteamfacilitatinglargelanguage}: This method employs LlamaIndex~\cite{llamaindexweb} for document retrieval and integrates a coding agent that performs debugging through interaction with a compiler environment for correcting runtime errors.
    \item \textsc{TFIDF+CodeAgent}: This uses TF-IDF indexing~\cite{salton1988term} for document retrieval and adopts the same coding agent as our method for code generation and debugging.
    \item \textsc{SentBERT+CodeAgent}: This performs retrieval using Sentence-BERT embeddings~\cite{reimers2019sentence} and likewise employs the same coding agent as our method for code generation and debugging.
\end{itemize}

\paragraph{(v) Human-involved code-based reasoning}
This category includes existing methods that rely on explicit human involvement during inference or prompt construction. Owing to their dependence on human-provided guidance, we refer to them as \emph{human-involved code-based reasoning} methods. Representative approaches in this category include \textsc{CodeGraph}~\cite{cai2024codegraphenhancinggraphreasoning} and \textsc{PIE}~\cite{gong2025pseudocodeinjectionmagicenablingllms}. Specifically,
\begin{itemize}[nosep, leftmargin=*]
    \item \textsc{CodeGraph}~\cite{cai2024codegraphenhancinggraphreasoning}: This method relies on human-selected, task-specific input–output code exemplars to guide graph reasoning, without a debugging mechanism. We provide the ground-truth functions in the technical documentation to support this setup.
    \item \textsc{PIE}~\cite{gong2025pseudocodeinjectionmagicenablingllms}: This approach assumes that LLMs are proficient in code generation, and therefore supplies task-specific pseudocode along with human-provided test cases to facilitate debugging.
\end{itemize}

\subsection{Detailed Related Work}
\subsubsection{LLM-based Graph Reasoning}
\label{sec:appendix_related_reasoning}
\noindent\textbf{Benchmarking LLMs on Graph Reasoning.}
Despite the remarkable success of LLMs in natural language processing tasks, their capabilities on graph-structured problems are important, particularly in the context of artificial general intelligence. A substantial body of prior work has investigated this challenge from the perspective of \emph{LLM-based graph learning}, which focuses on learning parameterized mapping functions over graph-structured data, from structural features to labels, to minimize expected risk and achieve generalization to unseen graph samples. Such methods typically target classical graph learning tasks, including node classification \cite{chen2024labelfree, tang2024graphgpt}, knowledge graph reasoning \cite{luo2024reasoning}, and molecular learning \cite{le2024molx}. Under this paradigm, LLMs primarily rely on capturing statistical patterns from data distributions, rather than performing explicit structural reasoning over graphs.
In contrast, \emph{LLM-based graph reasoning} aims to produce correct solutions through multi-step reasoning processes carried out by LLMs themselves over graph algorithm tasks, such as shortest paths or cycle check. The correct solutions to such tasks can be derived from a sequence of explicit graph algorithmic operations. 
Several studies refer to these problems as \emph{graph computation tasks} \cite{chen2024graphwizinstructionfollowinglanguagemodel, zhang2024gcoderimprovinglargelanguage, tang2025grapharena} to emphasize their well-defined algorithmic and executable semantics, whereas we adopt the term \emph{graph reasoning tasks} to highlight the role of LLMs in directly performing structural reasoning and decision-making over graph structures. Early benchmarks such as GraphQA \cite{fatemi2023talklikegraphencoding} and NLGraph \cite{wang2024languagemodelssolvegraph} were among the first to evaluate LLMs on graph reasoning tasks, including edge existence, node degree computation, cycle detection, shortest path, and tasks partially computable by graph neural networks. Their empirical results demonstrate that, while LLMs exhibit preliminary graph reasoning capabilities, their performance still falls short of ideal levels.
Motivated by these observations, subsequent work has proposed more systematic and comprehensive benchmarks and datasets for graph reasoning, including GraphWiz \cite{chen2024graphwizinstructionfollowinglanguagemodel}, GraphInstruct \cite{luo2024graphinstruct}, GraphWild \cite{zhang2024gcoderimprovinglargelanguage}, GTools \cite{wang2025GraphToolInstruction}, GraphArena \cite{tang2025grapharena}, and LLM4DyG \cite{zhang2024llm4dyg}. Among them, GraphWiz evaluates graph reasoning problems of varying complexity, including NP-complete tasks; GraphInstruct comprises $21$ graph analysis tasks spanning node-level, node-pair-level, and graph-level reasoning; and LLM4DyG focuses on dynamic graph reasoning involving temporal and spatial information.
However, existing benchmarks are largely restricted to small-scale graphs and single canonical tasks, and do not systematically evaluate LLMs’ ability to perform large-scale graph reasoning when graph sizes exceed the context window, nor complex reasoning that composes multiple graph tasks. To address this gap, we introduce a new graph reasoning dataset spanning multiple graph scales, which includes both simple and composite graph reasoning tasks. This dataset enables a more comprehensive assessment of LLMs in large-scale and complex graph reasoning.

\noindent\textbf{LLM-based Graph Reasoning Methods.}
Due to the unsatisfactory performance of LLMs on graph reasoning tasks, a growing body of recent work has sought to enhance their graph reasoning capabilities \cite{tang2024graphgpt, perozzi2024let, chen2024graphwizinstructionfollowinglanguagemodel, cai2024codegraphenhancinggraphreasoning, zhang2024gcoderimprovinglargelanguage, wang2025GraphToolInstruction, li2025graphteamfacilitatinglargelanguage, luo2024graphinstruct, chai2025graphllm, gong2025pseudocodeinjectionmagicenablingllms}. From the perspective of output reasoning execution, existing approaches can be broadly categorized into two categories: text-based graph reasoning, which solves graph reasoning tasks through natural language reasoning over graph structures serialized in textual form within the prompt; and code-based graph reasoning, which generates code and executes it to realize graph algorithmic computation.
\textit{(1) Text-based graph reasoning.}
This paradigm relies on LLMs’ ability to analyze text-serialized graph structures and typically employs various prompting strategies for natural language reasoning. For example, Talk-like-a-Graph \cite{fatemi2023talklikegraphencoding} and NLGraph \cite{wang2024languagemodelssolvegraph} systematically evaluate the effectiveness of standard textual instructions under different prompting settings, including zero-shot, few-shot, chain-of-thought \cite{wei2022chain}, and self-consistency \cite{wang2023selfconsistency}, for graph reasoning tasks. NLGraph further introduces Build-a-Graph, which explicitly requires LLMs to output a conceptual representation of the graph to enhance structural awareness, as well as Algorithmic Prompting, which provides algorithmic hints within prompts to facilitate reasoning. However, such purely text-based prompting approaches are constrained by the limited capacity of LLMs for deep algorithmic understanding and structured derivation, often resulting in suboptimal performance. To mitigate these limitations, several works incorporate instruction tuning to enhance graph reasoning ability. For instance, GraphWiz \cite{chen2024graphwizinstructionfollowinglanguagemodel} leverages instruction data optimized via direct preference optimization to guide LLMs toward more appropriate reasoning paths for graph computation tasks; GraphInstruct \cite{luo2024graphinstruct} proposes GraphLM, which applies LoRA fine-tuning with final-answer prediction as the optimization objective to improve graph understanding, and further introduces GraphLM+, which jointly optimizes key tokens in multi-step reasoning and the final answer to enhance graph reasoning performance; GraphLLM \cite{chai2025graphllm} compresses graphs into structured continuous vector prefixes aligned with the LLM representation space through fine-tuning, enabling the model to access a global graph summary at each attention layer. Nevertheless, prior studies indicate that instruction tuning exhibits limited transferability in graph reasoning tasks \cite{zhang2024can}, making robust generalization across different graph tasks challenging.
\textit{(2) Code-based graph reasoning.}
This paradigm equips LLMs with graph algorithmic capabilities through code generation and execution, without requiring explicit step-by-step natural language reasoning. For example, CodeGraph \cite{cai2024codegraphenhancinggraphreasoning} generates graph algorithm code using a small set of exemplars and obtains answers by executing the code with a program interpreter; GCoder \cite{zhang2024gcoderimprovinglargelanguage} further retrieves relevant implementation templates from a code repository based on embedding similarity to improve generalization; GraphTeam \cite{li2025graphteamfacilitatinglargelanguage} employs a multi-agent collaboration framework, in which a search agent retrieves relevant documents and prior experience via semantic retrieval, and a coding agent is responsible for code generation and compilation verification. However, both these text-based and code-based approaches still require serializing graph data into prompts. As the graph size grows beyond the LLM’s context window, these methods become increasingly infeasible, thereby severely limiting their effectiveness in large-scale graph reasoning scenarios.
To overcome the context window bottleneck, a small number of recent studies have begun to explore a \emph{task--graph decoupling paradigm}, in which prompts contain only task descriptions while graph data is processed externally through generated code. GraphTool-Instruction \cite{wang2025GraphToolInstruction} explicitly includes a complete set of graph tool descriptions in the prompt to guide LLMs in generating tool-invoking code; however, this approach relies on manually constructed tool sets and requires enumerating all tool descriptions within the prompt, limiting its scalability to large tool collections. PIE \cite{gong2025pseudocodeinjectionmagicenablingllms} guides code generation by injecting expert-level pseudocode derived from recent research papers and uses test cases to debug and select the optimal implementation, but it heavily depends on human-provided task-specific pseudocode, which restricts its practicality.
Based on these, we propose a code-based reasoning paradigm that decouples task descriptions from graph data and introduces a retrieval-augmented LLM-based agent framework to enable scalable graph reasoning beyond context window and human-involving constraints.

\subsubsection{Agents in RAG}
\label{sec:appendix_related_agentic_rag}
Autonomous agents are introduced by agentic RAG to enable dynamical retrieval and iterative refinement, which eliminates the constraint of static workflows in regular RAG. Specifically, an agentic RAG system utilizes an agent to decompose tasks, adaptively retrieve external knowledge, synthesize intermediate results, and iteratively performs refinement until a valid result is obtained.
For example, \citeauthor{ravuru2024agentic} propose to use agentic RAG for time series analysis with a hierarchical-agent framework.
CRAG~\cite{yan2024corrective} utilizes an evaluator to assess the quality of retrieved resources and perform corresponding actions based on the results. 
Considering the constraint-rich nature of structured documentation which requires explicit retrieval planning and iterative refinement, some work has proposed to utilize agentic RAG on structured documentation to enhance the reasoning and retrieval accuracy.
For instance, GeAR~\cite{shen2025gear} integrates base retrievals with a graph expansion mechanism and proposes a multi-step retrieval framework.
Similarly, HM-RAG~\cite{liu2025hm} provides parallel retrieval results from vector, graph, and web-based documents and integrates these results to improve answer accuracy. 
Recent work has adapted agentic RAG to graph reasoning, including GraphTeam~\cite{li2025graphteamfacilitatinglargelanguage}. They propose a code-based approach to solve graph reasoning problems with knowledge retrieval from external documentation.
However, they do not consider the hierarchical structure of technical documentation. On the other hand, our proposed framework bridges this gap by utilizing a layer-wise agentic retrieval framework for efficient localization. 


\begin{table*}[t]
\centering
\small
\setlength{\tabcolsep}{6pt}
\renewcommand{\arraystretch}{1.15}
\caption{Overview of graph reasoning tasks in the small-scale \texttt{\dataset-S}.}
\label{tab:small_scale_task_overview}
\resizebox{0.8\linewidth}{!}{
\begin{tabular}{lllll}
\toprule
\textbf{Task} & \textbf{Task Type} & \textbf{\#Examples} & \textbf{\#Nodes} & \textbf{\#Edges} \\
\midrule
Graph Connectivity Check (GCC) 
& Traversal-Based Reasoning 
& 674 
& 2--200 
& 1--19{,}701 \\

Connected Components Counting (CCC) 
& Traversal-Based Reasoning 
& 674 
& 2--200 
& 1--19{,}701 \\

Strongly Connected Components Counting (SCCC) 
& Traversal-Based Reasoning 
& 550 
& 4--200 
& 2--18{,}336 \\

Shortest Path (SP) 
& Traversal-Based Reasoning 
& 1{,}472 
& 2--200 
& 1--19{,}503 \\
\midrule
Eulerian Path Check (EPC) 
& Graph Property Detection 
& 1{,}224 
& 2--200 
& 1--18{,}336 \\

Graph Bipartiteness (GB) 
& Graph Property Detection 
& 1{,}224 
& 2--200 
& 1--18{,}336 \\

Graph Diameter (GD) 
& Graph Property Detection 
& 457 
& 2--200 
& 1--19{,}701 \\

Regularity Check (RC) 
& Graph Property Detection 
& 1{,}224 
& 2--200 
& 1--19{,}701 \\

Distance Regularity Check (DRC) 
& Graph Property Detection 
& 1{,}226 
& 3--200 
& 2--19{,}900 \\
\midrule
Maximum Flow (MF) 
& Combinatorial Optimization 
& 1{,}450 
& 2--200 
& 1--19{,}503 \\

Maximum Clique (MC) 
& Combinatorial Optimization 
& 674 
& 2--200 
& 1--19{,}701 \\

Maximum Independent Set (MIS) 
& Combinatorial Optimization 
& 674 
& 2--200 
& 1--19{,}701 \\
\midrule
Clustering Coefficient (CC) 
& Neighborhood / Structural Analysis 
& 1{,}117 
& 13--200 
& 14--19{,}701 \\

Common Neighbors (CN) 
& Neighborhood / Structural Analysis 
& 224 
& 16--200 
& 11--19{,}701 \\
\bottomrule
\end{tabular}
}
\end{table*}

\begin{table*}[h]
\centering
\small
\setlength{\tabcolsep}{6pt}
\renewcommand{\arraystretch}{1.15}
\caption{Overview of graph reasoning tasks in the large-scale \texttt{\dataset-L}.}
\label{tab:task_overview}
\resizebox{0.8\linewidth}{!}{
\begin{tabular}{lllll}
\toprule
\textbf{Task} & \textbf{Task Type} & \textbf{\#Examples} & \textbf{\#Nodes} & \textbf{\#Edges} \\
\midrule
Graph Connectivity Check (GCC) 
& Traversal-Based Reasoning 
& 45 
& 200--8{,}000 
& 5{,}400--301{,}158 \\

Connected Components Counting (CCC) 
& Traversal-Based Reasoning 
& 45 
& 200--8{,}000 
& 5{,}400--301{,}158 \\

Strongly Connected Components Counting (SCCC) 
& Traversal-Based Reasoning 
& 45 
& 200--8{,}000 
& 5{,}400--315{,}966 \\

Shortest Path (SP) 
& Traversal-Based Reasoning 
& 54 
& 200--8{,}000 
& 5{,}400--284{,}468 \\
\midrule
Eulerian Path Check (EPC) 
& Graph Property Detection 
& 90 
& 200--8{,}000 
& 5{,}400--315{,}966 \\

Graph Bipartiteness (GB) 
& Graph Property Detection 
& 90 
& 200--8{,}000 
& 5{,}400--315{,}966 \\

Graph Diameter (GD) 
& Graph Property Detection 
& 40 
& 200--8{,}000 
& 10{,}462--301{,}158 \\

Regularity Check (RC) 
& Graph Property Detection 
& 90 
& 200--8{,}000 
& 5{,}400--315{,}966 \\

Distance Regularity Check (DRC) 
& Graph Property Detection 
& 45 
& 200--8{,}000 
& 5{,}400--301{,}158 \\
\midrule
Maximum Flow (MF) 
& Combinatorial Optimization 
& 55 
& 200--8{,}000 
& 5{,}400--284{,}468 \\

Maximum Independent Set (MIS) 
& Combinatorial Optimization 
& 20 
& 200--1{,}800 
& 10{,}462--301{,}158 \\
\midrule
Clustering Coefficient (CC) 
& Neighborhood / Structural Analysis 
& 108 
& 200--8{,}000 
& 10{,}190--315{,}966 \\

Common Neighbors (CN) 
& Neighborhood / Structural Analysis 
& 17 
& 200--1{,}800 
& 12{,}807--301{,}158 \\
\bottomrule
\end{tabular}
}
\end{table*}

\begin{table*}[t]
\centering
\small
\caption{Overview of composite graph reasoning tasks in \texttt{\dataset-C}, including their algorithmic types, and example counts.
}
\label{tab:composite_tasks}
\resizebox{0.9\linewidth}{!}{%
\begin{tabular}{p{3.2cm} l c c c p{7.8cm}}
\toprule
\textbf{Task Name} & \textbf{Type} & \textbf{\#Examples} & \textbf{\#Nodes} & \textbf{\#Edges} & \textbf{Primitive Tasks} \\
\midrule

Clustering--Shortest Path &
Sequential 
& 572
& 2--200
& 1--19{,}701
& Shortest Path; Clustering Coefficient \\

SCC--Diameter &
Sequential 
& 550
& 4--200
& 2--18{,}336
&Strongly Connected Components; Graph Diameter \\

CC-Flow-Clustering &
Sequential 
& 94
& 2--200
& 1--19{,}701
& Connected Components; Maximum Flow; Clustering Coefficient \\

\midrule

Pair Tightness Score &
Parallel 
& 339
& 3--200
& 1--18{,}915
&Shortest Path; Common Neighbors; Clustering Coefficient \\

Bridge Hub Identification &
Parallel 
& 674
& 2--200
& 1--19{,}701
& Clustering Coefficient; Shortest Path \\

Endpoint-Aware Flow &
Parallel 
& 131
& 8--198
& 5--19{,}503
& Maximum Flow; Clustering Coefficient \\

\midrule

Eulerian Path--Diameter &
Conditional 
& 333
& 2--200
& 1--19{,}701
& Eulerian Path Check; Graph Diameter; Clustering Coefficient \\

Connectivity--Component Diameter &
Conditional 
& 674
& 2--200
& 1--19{,}701
& Graph Connectivity Check; Connected Components; Graph Diameter \\

SCC--Eulerian Check &
Conditional 
& 550
& 4--200
& 2--18{,}336
& Eulerian Path Check; Strongly Connected Components \\

\bottomrule
\end{tabular}%
}
\end{table*}

\begin{dataexample*}{{\dataset}-S and -L Data Example (Shortest Path)}
    \textbf{Task Description:}
    You are given a directed, weighted graph represented by an edge list, where each edge is a triple $(u, v, w)$ indicating a directed connection from node $u$ to node $v$ with weight $w$. Given a source node $\textit{source}$ and a target node $\textit{target}$, your task is to compute the shortest path from $\textit{source}$ to $\textit{target}$. The shortest path is defined as a path that minimizes the sum of edge weights among all possible directed paths connecting the two nodes in the graph. Return only the length of the shortest path as a floating-point value. 

    \textbf{Edge List:}
    $
    \{
    (3,4,5.7),
    (3,1,6.5),
    (3,6,5.5), 
    (3,5,5.2),
    (3,0,8.1),
    (3,2,2.6), 
    \ldots
    \}
    $
    
    \textbf{Arguments:} $\textit{source}=3$, $\textit{target}=5$
    
    \textbf{Answer:} $7.3$ 
\end{dataexample*}

\begin{dataexample*}{\dataset-C Data Example (SCC--Eulerian Check)}
    \textbf{Task Description}:
    You are given a directed, unweighted graph and are asked to compute a \emph{structural complexity score} based on two properties of the graph: the existence of an Eulerian path and the structure of its strongly connected components.
    The graph is provided as an edge list, which is a sequence of ordered pairs $(u, v)$, where each pair represents a directed edge from node $u$ to node $v$. An \emph{Eulerian path} is defined as a directed trail that traverses every edge of the graph exactly once. A \emph{strongly connected component} (SCC) is a maximal set of nodes such that for any two nodes $a$ and $b$ in the set, there exists a directed path from $a$ to $b$ and from $b$ to $a$.
    If the graph admits an Eulerian path, the structural complexity score is defined as the number of nodes in the largest strongly connected component. Otherwise, the score is defined as the total number of strongly connected components in the graph. Return the resulting structural complexity score as an integer.

    \textbf{Edge List:}
    $
    \{\, 
    (0,1),\,
    (0,2),\,
    (0,3),\,
    (0,4),\,
    \ldots
    \,\}
    $
    
    \textbf{Arguments:} None
    
    \textbf{Answer:}$32$ 
\end{dataexample*}

\end{document}